\documentclass[twocolumn,aps,prl,reprint,superscriptaddress]{revtex4-1}
\usepackage{amsmath}
\usepackage{amssymb}
\usepackage{graphicx}
\usepackage{color}
\usepackage{tabularx}
\usepackage{hyperref}
\usepackage[caption=false]{subfig}
\usepackage[colorinlistoftodos]{todonotes}
\usepackage{lipsum}
\usepackage{textcomp}
\usepackage{multibib}
\newcites{main}{Main text}
\newcites{meth}{Methods}
\newcites{sm}{Supplementary Materials}

\usepackage{tikz}
\usepackage{circuitikz}
\usepackage{tikz-timing}
\usetikzlibrary{arrows.meta,decorations.pathmorphing,decorations.pathreplacing,positioning,shapes,fadings}

\graphicspath{ {main_figures/} {supplement_figures/} }

\makeatletter
\@ifundefined{textcolor}{}
{%
 \definecolor{BLACK}{gray}{0}
 \definecolor{WHITE}{gray}{1}
 \definecolor{RED}{rgb}{1,0,0}
 \definecolor{GREEN}{rgb}{0,1,0}
 \definecolor{BLUE}{rgb}{0,0,1}
 \definecolor{CYAN}{cmyk}{1,0,0,0}
 \definecolor{MAGENTA}{cmyk}{0,1,0,0}
 \definecolor{YELLOW}{cmyk}{0,0,1,0}
}
\usepackage{dcolumn}
\usepackage{bm}
\makeatother

\begin{document}

\title{Spectroscopy of Wigner molecules on superfluid helium \\
using a superconducting resonator}

\author{G. Koolstra}
\email{gkoolstra@uchicago.edu}
\affiliation{The James Franck Institute and Department of Physics, University of Chicago, Chicago, Illinois 60637, USA}
\author{Ge Yang}
\affiliation{The James Franck Institute and Department of Physics, University of Chicago, Chicago, Illinois 60637, USA}
\author{David I. Schuster}
\email{David.Schuster@uchicago.edu}
\affiliation{The James Franck Institute and Department of Physics, University of Chicago, Chicago, Illinois 60637, USA}
\date{\today}
\maketitle

\textbf{Electrons on helium form a unique two-dimensional electron system on the interface of liquid helium and vacuum \citemain{MonarkhaKono2004}. On liquid helium, trapped electrons can arrange into strongly correlated states known as Wigner molecules \citemain{RousseauPRB2009}, which can be used to study electron interactions in the absence of disorder, or as a highly promising resource for quantum computation \citemain{Platzman1999, DykmanPRB2003, Lyon2006, Schuster2010}. Wigner molecules have orbital frequencies in the microwave regime and can therefore be integrated with circuit quantum electrodynamics (cQED), which studies light-matter interactions using microwave photons \citemain{WallraffNature2004}. Here, we experimentally realize a cQED platform with the orbital state of Wigner molecules on helium. We deterministically prepare one to four-electron Wigner molecules on top of a microwave resonator, which allows us to observe their unique spectra for the first time. Furthermore, we find a single-electron-photon coupling strength of $g/2\pi = 4.8\pm0.3$ MHz, greatly exceeding the resonator linewidth $\kappa/2\pi = 0.5$ MHz. These results pave the way towards microwave studies of strongly correlated electron states and coherent control of the orbital and spin state of Wigner molecules on helium.}

The orbital state of electrons on helium consists of the lateral motion of strongly correlated electrons. Since the electron-phonon coupling in helium is small compared with semiconductors, this motion is expected to have low dissipation, making the orbital state an attractive candidate for a long-lived quantum bit \citemain{Schuster2010, DaniilidisNJOP2013}. In addition, by adding a magnetic field gradient from a micro-magnet \citemain{PioroNatPhys2008}, the orbital state offers a path towards the electron spin state \citemain{Viennot2015, Mi2018, LandigNature2018, Samkharadze2018Science, PengPRA2017}. Since the orbital frequency of electrons on helium is in the microwave regime, and electrons can couple strongly to microwave photons \citemain{Schuster2010, ShlomiPRA2017, Mi2017, Stockklauser2017}, cQED can play a unique role in the detection and manipulation of the orbital state.

A small ensemble of electrons on helium behaves differently from other confined electron systems, such as semiconductors or atoms, where the electron wavefunctions are delocalized and overlap. On the surface of liquid helium electron interactions dominate \citemain{Rees2011PRL, ReesPRL2012} and are largely unscreened, which results in strongly correlated electron configurations known as Wigner molecules. The symmetry of these molecules changes for each additional electron, which has been observed in charging diagrams of small islands of liquid helium \citemain{Papageorgiou2005, RousseauPRB2009}. Additionally, theory has predicted Wigner molecule configurations and orbital frequencies in various trapping potentials \citemain{BoltonSandM1993, BedanovPRB1994, SchweigertPRB1995, KongPRE2002}. Spectroscopy of Wigner molecules on helium could provide insight into both the internal molecular structure and the molecule's environment, but the lack of a microwave interface has prevented this to date.

Here we realize the coupling of Wigner molecules on helium to a microwave cavity that serves as an electron detector and harbors an electron reservoir. We transfer electrons from the reservoir to a small island where we control the charge with single electron resolution and perform spectroscopy of a single electron and few-electron Wigner molecules. We observe unique spectra which serve as a fingerprint for the molecule's internal structure, and a large electron-photon coupling. These results open the door to coherent control of the orbital and spin state of Wigner molecules on helium.

\begin{figure}[t]
\includegraphics[width=\columnwidth]{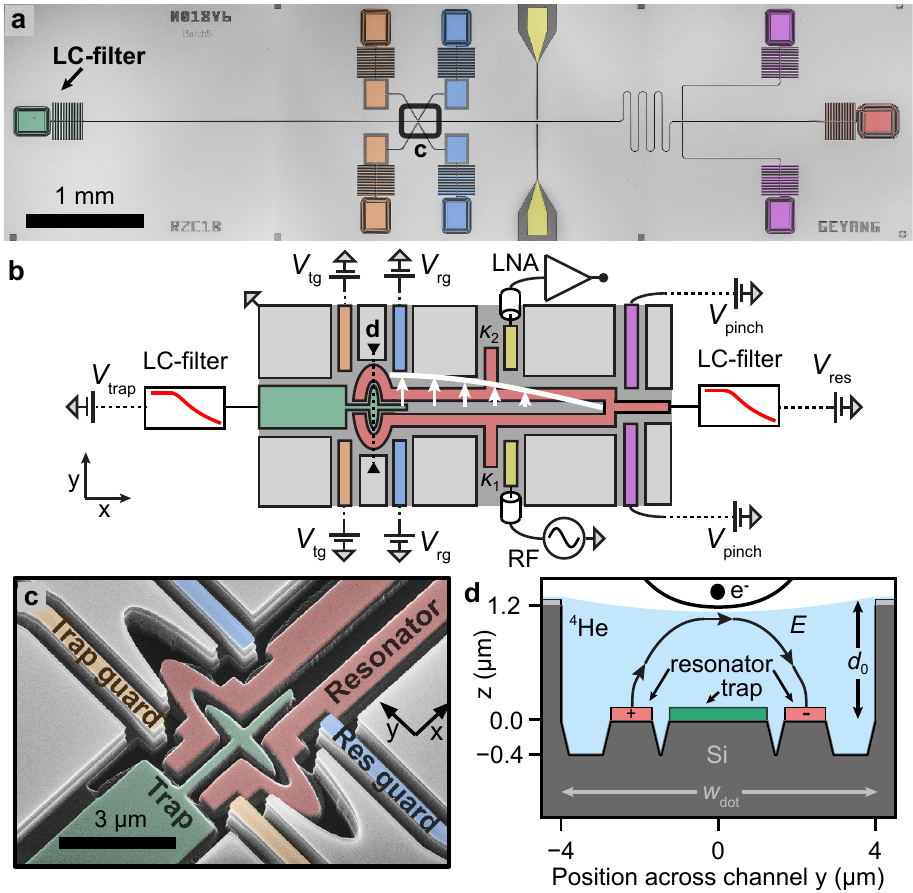}
\caption{\textbf{An electron-on-helium dot} \textbf{(a)} Optical micrograph and \textbf{(b)} schematic of the device. The resonator (red) can be probed with an RF tone via coplanar waveguides (yellow) that couple (decay rates $\kappa_{1,2}$) to the microwave resonator. The white arrows show the electric field of the $\lambda/4$ microwave mode at the center of the channel. The transmission is amplified with a low-noise amplifier (LNA). The electrostatic potential for electrons is controlled with additional electrodes, which are all equipped with individual low-pass filters to reject noise at the resonance frequency \protect\citemain{Mi2016APL} . In (b), we only show these filters for the trap and resonator. \textbf{(c)} Tilted, false-colored scanning electron micrograph of the dot showing the micro-machined silicon substrate. The resonator (red) and trap electrode (green) are located on the bottom of a micro-channel, which lies 1.2 \textmu{}m below the level of the resonator guards (blue), trap guards (orange) and ground plane. \textbf{(d)} Schematic cross-section of the dot shown in (c), depicting the resonator center pins and trap electrode submersed in liquid helium. Wigner molecules are trapped on the interface of liquid $^4$He and vacuum by the electrostatic potential (solid black line) generated by electrodes near the dot. The electron orbital state couples to the transverse microwave electric field $E$ from the resonator.}
\label{fig:fig1}
\end{figure}

At the heart of our cQED device lies a superconducting microwave resonator with an integrated electron-on-helium quantum dot (Fig.~\ref{fig:fig1}a). Our coplanar stripline resonator consists of two niobium center pins, which are joined at one end (Fig.~\ref{fig:fig1}b, c) and are situated below the ground plane at the bottom of a micro-channel (width $w = 3.5$ \textmu{}m, and depth $d_0 \approx $ 1.2 \textmu{}m). The microwave mode with resonance frequency $f_0 = 6.399$ GHz and linewidth $\kappa_\mathrm{tot}/2\pi=0.4$ MHz has an RF electric field that is concentrated between the center pins. As liquid $^4$He fills the channel, its surface is stabilized due to helium's surface tension, after which the liquid helium can serve as a defect-free substrate for electrons (Fig.~\ref{fig:fig1}d). 

\begin{figure}[t]
\includegraphics[width=\columnwidth]{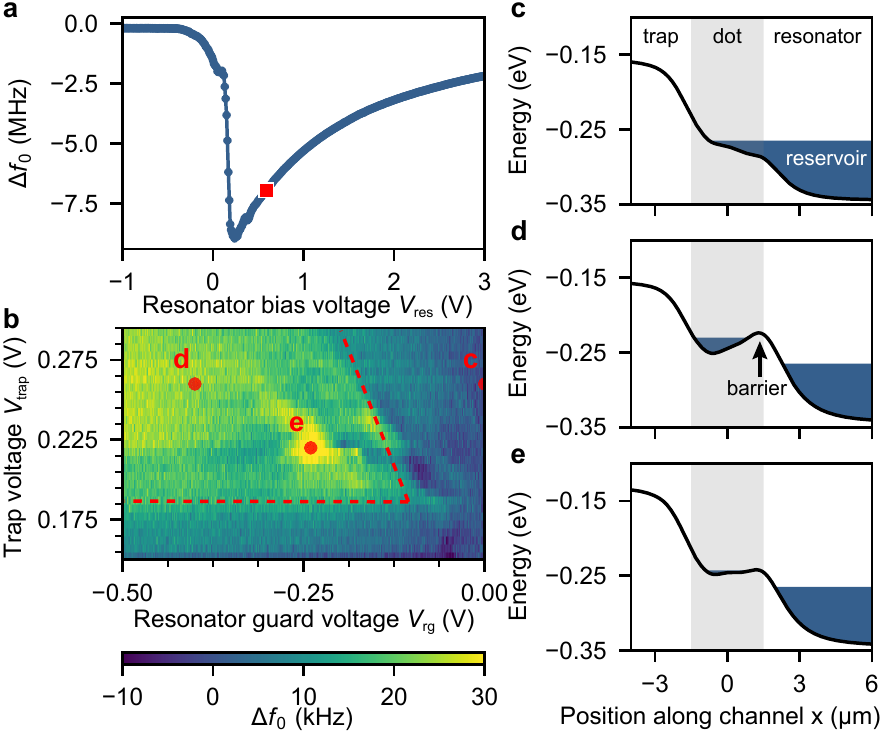}
\caption{\textbf{Separating electrons from the reservoir} \textbf{(a)} At $T = 25$ mK reservoir electrons are detected through a dispersive resonance frequency shift which depends on $V_\mathrm{res}$. The jump in $\Delta f_0$ at $V_\mathrm{res} \approx 0.2$ V is consistent with electron loss from an ensemble with density $n \approx 6 \times 10^{12}$ m$^{-2}$. The data presented hereafter are taken with the resonator bias voltage fixed at 0.6 V, which is marked by a square.
\textbf{(b)} Measured resonance frequency shift while raising a barrier between the dot and reservoir as function of $V_\mathrm{trap}$. The dashed line segments mark the border of a region where electrons can be trapped in the dot. The largest $\Delta f_0$ are expected when the electron orbital frequency approaches $f_0$. For $V_\mathrm{trap} > 0.3$ V electron trapping is unstable, because reservoir electrons can freely flow through the dot onto the trap electrode. \textbf{(c)-(e)} Simulated potential energy along the channel for three different values of $V_\mathrm{rg}$, $V_\mathrm{trap}$, marked by the red dots in (b). Reservoir electrons ($x>2$ \textmu{}m) and electrons in the dot ($-1.5$ \textmu{}m $< x < 1.5$ \textmu{}m) are represented as a constant energy (blue). Electrons are trapped in the dot in (d) and (e).}
\label{fig:fig2}
\end{figure}

After depositing electrons over the resonator, we detect a dispersive resonance frequency shift that depends strongly on the resonator bias voltage $V_\mathrm{res}$ (Fig.~\ref{fig:fig2}a) and the number of electrons on the resonator \citemain{GeYang2016}. For the experiments presented hereafter, we fix $V_\mathrm{res}$ at 0.6 V such that electrons on the resonator can be treated as a reservoir with constant electron density. Furthermore, our measurements are performed at $T = 25$ mK and low incident microwave power ($n_\mathrm{ph} \approx 5$) such that electrons respond linearly to the resonator's driving force.

\begin{figure*}[t]
\includegraphics[width=\textwidth]{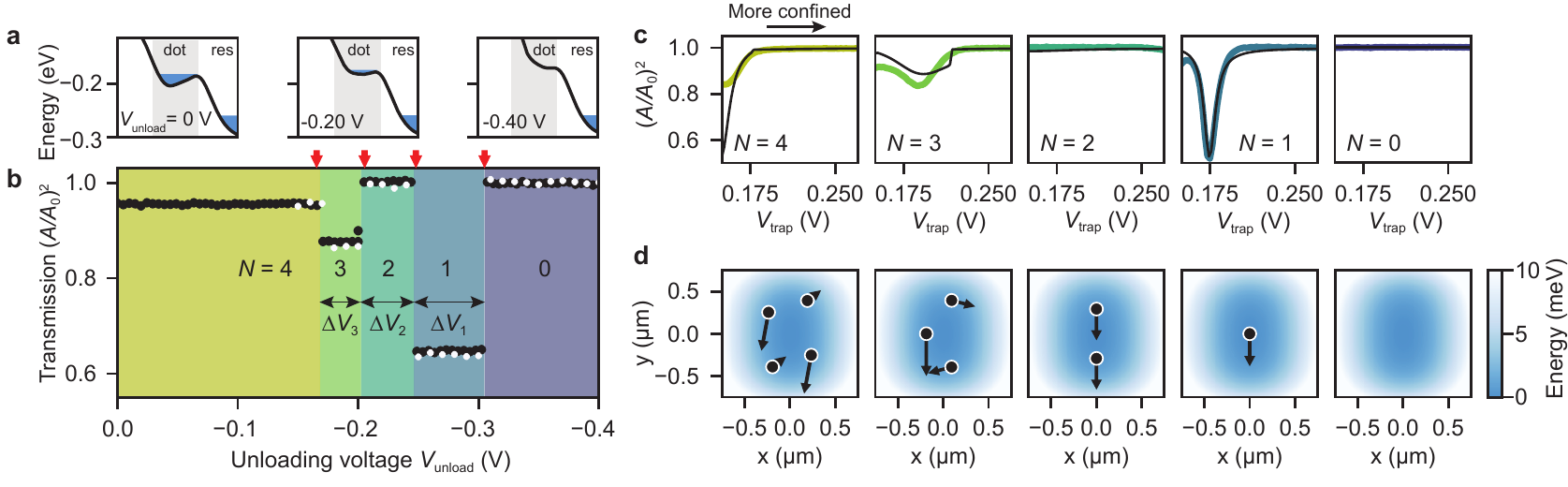}
\caption{\textbf{Resonator signatures of few-electron Wigner molecules} \textbf{(a)} Schematic of the unloading procedure. At the unloading voltage, the dot's trap depth decreases for more negative $V_\mathrm{unload}$. No electrons can occupy the dot at $V_\mathrm{trap} = -0.4$V. \textbf{(b)} With decreasing $V_\mathrm{unload}$, sudden changes in the resonator transmission (black dots, measured at $V_\mathrm{trap} = 0.175$ V and $V_\mathrm{tg} = 0$ V) indicate that electrons leave the dot. We observe five distinct plateaus that are reproduced after reloading the dot eight hours later (white dots), and are associated with a constant number of trapped electrons $N$. Red arrows indicate predicted escape voltages for $N=4$ to 1 electrons (left to right) from a single-parameter model, see Supplementary Table I. \textbf{(c)} Spectra of Wigner molecules consisting of up to four electrons, measured by varying the trap curvature using $V_\mathrm{trap}$. Below $V_\mathrm{trap} = 0.15$ V electron trapping is unstable. The solid black lines are simulated cavity responses (see Methods) and agree qualitatively with the measured spectra. The discontinuity in the simulation for $N = 3$ is due to a sudden change in position of the electrons, and is not expected to be visible in the averaged data. \textbf{(d)} Simulated electron configurations in the approximated electrostatic potential, shown for $V_\mathrm{trap}=0.175$ V. The arrows show the electron motion for the eigenmode that is most strongly coupled to the resonator. The microwave electric field is in the $y$ direction.}
\label{fig:fig3}
\end{figure*}

We use the dot in Fig.~\ref{fig:fig1}c to isolate individual electrons from the reservoir, which requires fine control over the electrostatic potential. We achieve this using three sets of electrodes near the tip of the resonator where the microwave electric field is strongest. The size of the electrodes near the dot is much larger than in semiconducting quantum dots, because the unscreened electron interaction results in inter-electron distances exceeding 200 nm. With appropriate voltages applied to the electrodes, the smooth electrostatic potential (Fig.~\ref{fig:fig2}d,e) allows for trapping of electrons. Furthermore, due to the dot's oblong shape, the lateral motion of trapped electrons is primarily in the $y$-direction (see Fig.~\ref{fig:fig1}d), such that it couples to the transverse microwave field of the resonator.

To load the dot we use the trap electrode (Fig.~\ref{fig:fig1}c, green) to attract reservoir electrons towards the dot, and the resonator guard (blue) to create a barrier between the dot and reservoir. Only if the trap voltage is sufficiently positive, and the resonator guard is sufficiently negative can electrons be loaded and contained in the dot, respectively. When monitoring the resonance frequency shift $\Delta f_0$ in response to these two voltages, we only see significant signal in an area that is marked by two converging dashed lines in Fig.~\ref{fig:fig2}b. The dashed lines are obtained from simulation of the electrostatic potential near the dot (see Methods), and indicate the presence of a barrier between reservoir electrons and electrons in the dot. Well within the predicted trapping region, we observe resonance frequency shifts that depend sensitively on $V_\mathrm{trap}$ and $V_\mathrm{rg}$, indicating that trapped electrons in the dot interact with the resonator. The observed shift depends on the number of trapped electrons, which increases for a larger trap voltage, as well as the shape of the electrostatic potential.

To deterministically prepare few-electron Wigner molecules, we partially unload the dot using the trap guard electrode (orange in Fig.~\ref{fig:fig1}c). A partial unload consists of briefly sweeping the trap guard voltage to $V_\mathrm{unload} < 0$, which decreases the trap depth (see Fig.~\ref{fig:fig3}a), followed by a measurement of the resonator transmission at $(V_\mathrm{trap}, V_\mathrm{tg}) = (0.175, 0.0)$ V. The plateaus in resonator transmission shown in Fig.~\ref{fig:fig3}b are reproduced after reloading the dot, but are absent when the dot is initially empty. Therefore, each plateau is associated with a constant number of trapped electrons, and the final change in transmission at $V_\mathrm{unload} = -0.305$ V leaves the dot empty.

The sudden changes in transmission are consistent with single electrons leaving the dot. We show this by modeling the trap as an axially symmetric harmonic well in which the electron configurations can be calculated analytically \citemain{BedanovPRB1994, SchweigertPRB1995}. From the voltage at which the last electron escapes, we estimate unloading voltages for two, three and four electrons, using the effective trap curvature as the only free parameter (see Methods). Red arrows in Fig.~\ref{fig:fig3}b indicate these estimates, and agree within 3 mV with the plateau edges. This unloading method therefore allows for deterministic preparation of one to four-electron Wigner molecules.

\begin{figure*}[t]
\includegraphics[width=\textwidth]{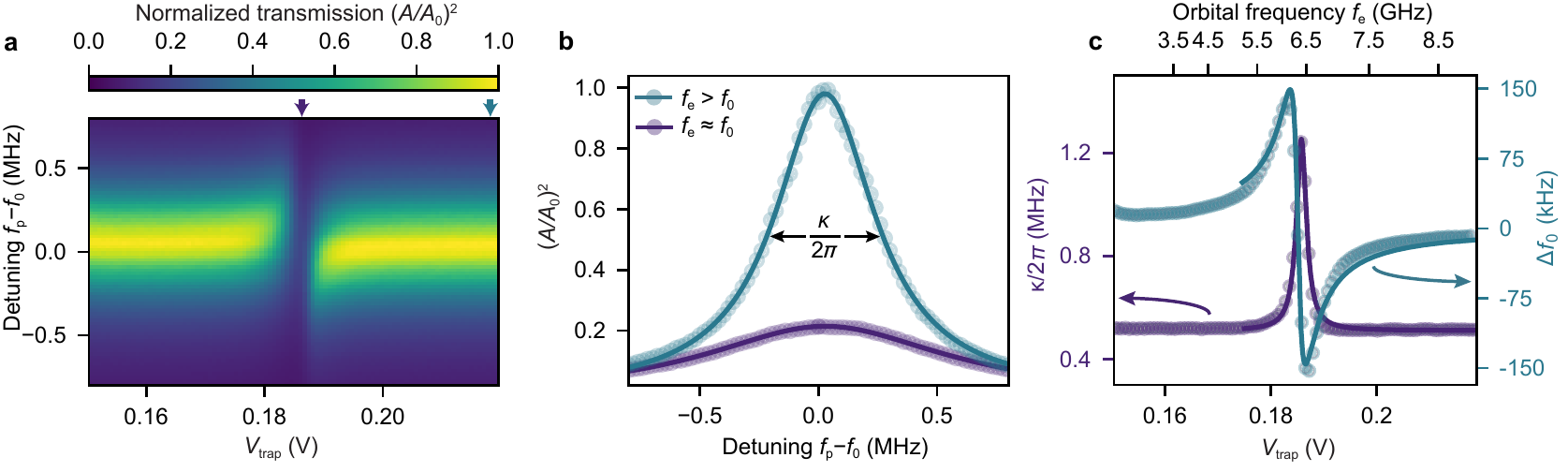}
\caption{\textbf{Single electron resonator spectroscopy} \textbf{(a)} Normalized transmission amplitude as function of trap voltage and microwave probe detuning $f_p - f_0$. \textbf{(b)} Resonator spectra for two values of $V_\mathrm{trap}$, indicated by arrows on the horizontal axis in (a). For $V_\mathrm{trap}$ = 0.184 V (0.23 V) the electron is on (off) resonance with the cavity. The resonant trace illustrates the sensitivity of our device to a single electron. \textbf{(c)} Resonance frequency shift (right axis) and resonator decay rate (left axis) obtained by fitting the Lorentzian resonator spectra from (a). The solid line is a fit to a model that yields a coupling strength near resonance of $g/2\pi = 4.8 \pm 0.3$ MHz and total electron linewidth $\gamma/2\pi = 77\pm19$ MHz. The top horizontal axis displays how the electron orbital frequency varies as function of $V_\mathrm{trap}$, and shows a crossing with the resonator ($f_e = 6.4$ GHz) at $V_\mathrm{trap}$ = 0.184 V.}
\label{fig:fig4}
\end{figure*}

The increasing length of transmission plateaus $\Delta V_N$ with decreasing $N$ is a telltale sign of strongly interacting electron ensembles, such as Wigner molecules \citemain{RousseauPRB2009}, and is in stark contrast to an equally-spaced charging diagram typically seen in metallic islands or semiconducting quantum dots. For an electron on helium dot, an unscreened interaction results in a significantly different charge configuration each time an electron is removed from the dot (Fig.~\ref{fig:fig3}d), resulting in the characteristic irregular spacing.

While a Wigner molecule is trapped in the dot, we use the resonator to observe it's unique spectrum, which provides insight in the electron configurations and orbital frequencies. We perform spectroscopy by monitoring the resonator's transmission while varying the trap voltage, which deforms the trap and therefore controls the orbital frequencies. For this measurement, a Wigner molecule can be trapped and studied for hours, since the trap depth is large compared to the zero-point energy and thermal energy. Fig.~\ref{fig:fig3}c shows five different spectroscopy traces, each corresponding to the different-sized Wigner molecules from Fig.~\ref{fig:fig3}b. To retrieve electron configurations and orbital frequencies, we numerically minimize the total energy of the ensemble and solve the coupled equations of motion \citemain{GeYang2016}. The electron configurations (Fig.~\ref{fig:fig3}d) change significantly as electrons are added or removed from the dot, and show correlated electron motion, originating from strong electron interactions. The largest signal in Fig.~\ref{fig:fig3}c occurs for a single electron at $V_\mathrm{trap}=0.175$ V when its orbital frequency is resonant with the resonator. In our model, the orbital frequency of larger Wigner molecules remains detuned for all $V_\mathrm{trap}$, which is due to a strong anharmonic component in the electrostatic potential. From the quartic term in this potential, we estimate a single-electron anharmonicity of 85 MHz, which holds promise for creating an electron-on-helium orbital state qubit.

We now focus on a single trapped electron and investigate its properties by tuning the orbital frequency into resonance with the resonator. Fig.~\ref{fig:fig4}a shows a crossing of the orbital frequency with the resonator around $V_\mathrm{trap} = 0.184$ V, which is accompanied by a rapid change in $\Delta f_0$ (Fig.~\ref{fig:fig4}c). By fitting the measured frequency shift to a model, which takes into account one orbital mode coupled to a single resonator mode \citemain{Cottet2017}, we obtain a single-electron-photon coupling strength $g = 2\pi \times (4.8 \pm 0.3)$ MHz and electron linewidth $\gamma = \gamma_1/2 + \gamma_\varphi = 2\pi \times (77 \pm 19)$ MHz. The coupling strength is large compared to the resonator linewidth ($\kappa/2\pi \approx 0.5$ MHz), indicating that each photon measures the presence of the electron, and the coupling is similar to that measured in semiconducting quantum dot cQED architectures \citemain{Mi2017}. In addition, our estimate of the anharmonicity (see Supplementary Figure 7) is similar to that in superconducting qubits, indicating that with a reduced linewidth the orbital state of a single electron on helium can be used as a qubit.

The total linewidth $\gamma$ is three orders of magnitude larger than expected from the electron-phonon coupling in $^4$He and charge noise from the bias electrodes, respectively ($\gamma/2\pi < 0.1$ MHz) \citemain{Schuster2010}. We identify the dominant source of excess noise as classical helium fluctuations in the dot, caused by the pulse tube refrigerator. This is corroborated by a measurement of the crossing voltage as function of time, which shows spectral features of the pulse tube refrigerator. To estimate the dephasing rate due to helium fluctuations, we estimate an electron's sensitivity to helium fluctuations from electrostatic simulations ($\partial f_e / \partial t_\mathrm{He} \approx 80$ MHz/nm) and independently measure helium fluctuations ($\Delta t_\mathrm{He} \approx 1.4$ nm), yielding $\gamma_\varphi/2\pi \approx 110$ MHz. Therefore, we expect the single electron linewidth to be limited by dephasing due to helium level fluctuations.

Reducing the linewidth and increasing the coupling strength offers a path towards coherent control of a single electron on helium and may enable more accurate spectroscopic studies of Wigner molecules, through direct measurement of the electron orbital frequencies using two-tone spectroscopy \citemain{SchusterPRL2005}. In the next generation of electron on helium dots, one can passively or actively reduce the vibrations that excite the helium surface \citemain{PelliccioneRSI2013}, or engineer a dot geometry that has a reduced sensitivity to classical helium vibrations. In addition, microwave resonators made of high kinetic inductance superconductors can enhance the coupling strength in our device via an increased characteristic impedance \citemain{Samkharadze2018Science, Shearrow2018}.

In conclusion, we have integrated an electron-on-helium dot with a superconducting microwave resonator and observed distinct spectra of Wigner molecules consisting of up to four electrons. The large anharmonicity and coupling strength of a single electron on helium hold promise for creating an electron-on-helium qubit, which can be readily integrated with superconducting qubits while leveraging established protocols. Finally, when combined with a magnetic field gradient, the orbital state offers a clear path towards control of single electron and Wigner molecule spin states.


\section*{Acknowledgements}
\begin{acknowledgments} 
This research was supported by the DOE, Office of Basic Energy Sciences, Materials Sciences and Engineering Division. We thank K.W. Lehnert for the parametric amplifier used in this work, S. Chakram, D.D. Awschalom, D.J. van Woerkom, E. Kawakami, and other members of the Schuster lab for insightful discussions and P.J. Duda and D.C. Czaplewski for assistance and advice during fabrication of the device. This work made use of the Pritzker Nanofabrication Facility of the Institute for Molecular Engineering at the University of Chicago, which receives support from Soft and Hybrid Nanotechnology Experimental (SHyNE) Resource (NSF ECCS-1542205), a node of the National Science Foundation’s National Nanotechnology Coordinated Infrastructure.
\end{acknowledgments} 

\section*{Author contributions}
The experiment was conceived by D.I.S. All authors contributed to the design of the experiment. G.Y. and G.K. fabricated the samples. G.K. performed the measurements and analyzed the data. G.K. and D.I.S. wrote the manuscript and all authors commented on the manuscript. 

\section*{Competing interests}
The authors declare no competing interests.

\section*{Methods}
\noindent\textbf{Fabrication.} First an 80 nm thick Nb ground plane was evaporated onto a high-resistivity ($>10$ k$\Omega$cm) Si $\langle 100 \rangle$ wafer, followed by deposition of a 100 nm thick silicon oxide sacrificial layer, which was used to protect the Nb ground plane during the following etch steps. The micro-channels were defined using a Raith EBPG-5000+ electron beam lithography system and etched using a CHF$_3$/SF$_6$ chemistry, immediately followed by an HBr/O$_2$ etch. In the second step the resonator center pins were defined using e-beam lithography. After development, evaporation of a 150 nm thick Nb layer and lift-off, the center pins remained on the bottom of the micro-channel. To improve robustness of the device and avoid electrical breakdown at low temperatures, we etched away an additional $\sim$400 nm of Si substrate in between the resonator center pins. To this end, another layer of 80 nm thick silicon oxide was deposited, after which the additional Si was etched with the previously described etch chemistry. The silicon oxide layer was removed using buffered HF and a DI water rinse.
\\
\\
\textbf{Measurements.} All measurements were performed in an Oxford Triton 200 dilution refrigerator with a base temperature of 25 mK. The chip was mounted in a custom-designed hermetic sample cell and sealed with indium to prevent superfluid helium leaks. Helium was supplied to the sample cell from a high purity $^4$He gas cylinder and, using a control volume ($V \approx 25$ cm$^3$) in a gas handling system, we were able to introduce a controlled amount of helium to the sample cell. The experiment was performed in a regime where the channel was almost full and the liquid helium film was stabilized due to surface tension \citemeth{MMartyJPC1986}.

Electrons were captured on the helium surface by thermal emission from a tungsten filament situated above the chip, while applying a positive voltage to the resonator DC bias electrode ($V_\mathrm{res} = 3.0$ V) and a negative bias voltage to the filament. We assume electrons in the reservoir were distributed uniformly across the resonator and estimate the electron density from the resonator voltage at which electrons can no longer be contained on the resonator, as depicted by the sudden increase in $\Delta f_0$ in Fig.~\ref{fig:fig2}a. At $V_\mathrm{res}^\mathrm{th} = 0.18$ V we estimate the density
\begin{equation}
    n \approx \frac{\varepsilon_0 \varepsilon_\mathrm{He}}{e t_\mathrm{He}} a_\mathrm{res} V_\mathrm{res}^\mathrm{th} = 6 \times 10^{12} \, \mathrm{m}^{-2}, \label{eq:density_determination}
    \
\end{equation}
where $a_\mathrm{res}$ is the resonator electrode lever arm, $t_\mathrm{He}$ is the helium thickness, $\varepsilon_\mathrm{He} = 1.056$ is the dielectric constant of helium and $e$ is the elementary charge. This density corresponds to approximately $10^5$ reservoir electrons, whose orbital frequency stayed far detuned from $f_0$ during experiments with electrons in the dot. 

The pulse tube refrigerator is a continuous source of mechanical vibrations which excites the liquid helium surface. These vibrations were detected by the microwave resonator as a slowly varying resonance frequency jitter, with a standard deviation of approximately $6.8$ kHz in the absence of reservoir electrons. This jitter complicated the measurement of small resonance frequency shifts due to trapped electrons, which were typically of the same order as the jitter. However, since the dominant frequency components in the mechanical noise spectrum were below 10 Hz, we circumvented this issue by sweeping electrode voltages faster than 1/10 Hz$^{-1}$, such that signatures of trapped electrons became visible after averaging.
\\
\\
\textbf{Electrostatic simulations of the dot.}
The electrostatic potential near the dot was obtained by solving Poisson's equation using the finite element method with \textsc{ansys maxwell}. We separately solve the potential for each electrode that contributes to the dot potential by applying 1 V on a single electrode while keeping all other electrodes grounded. We minimize numerical noise in the potential by increasing the vertex density in the center of the dot and imposing strong convergence criteria. For post-processing the potential values are cast to a regular Cartesian grid using interpolation. 

The two converging dashed line segments in Fig.~\ref{fig:fig2}b are obtained by considering both the potential along the channel and the reservoir density. The reservoir density $n$ sets the chemical potential of the reservoir via (approximately) $e^2 n t_\mathrm{He} / \varepsilon_0 \varepsilon_\mathrm{He}$, and for larger $n$, $V_\mathrm{rg}$ must be more negative to maintain a barrier between reservoir and dot (Fig.~\ref{fig:fig2}d). For our device, this non-zero barrier condition is captured by a line segment with slope 1.15. The reservoir density $n$ determines the offset of this line segment, and was measured by increasing $V_\mathrm{trap}$ until electron transport occurred onto the trap electrode. From an equation similar to Eq. \eqref{eq:density_determination} we find $n \approx 4 \times 10^{12}$ m$^{-2}$. The horizontal line segment was found by finding the minimum $V_\mathrm{trap}$ for which the reservoir extends left of $x = 1.5$ \textmu{}m at $V_\mathrm{rg} = 0$. Fig.~\ref{fig:fig2}c shows a situation above this threshold, for which the loading operation should result in trapped electrons.
\\
\\
\textbf{Unloading the dot.}
The dot was unloaded by sweeping the trap guard to $V_\mathrm{tg} = V_\mathrm{unload} < 0$ while keeping all other electrodes constant at $(V_\mathrm{res}, V_\mathrm{trap}, V_\mathrm{rg}) = (0.6, 0.15, -0.4)$ V. The electrodes were then ramped back to $(V_\mathrm{trap}, V_\mathrm{tg}) = (0.175, 0)$ V in order to probe the resonator transmission. The speed of the ramp did not change the charging diagram of Fig.~\ref{fig:fig3}b.

To confirm that changes between transmission plateaus in Fig.~\ref{fig:fig3}b are associated with single electron transport, we simulated unloading using a combination of electrostatic simulations and analytical calculations. Even though the electrode geometry in the dot produced a complex and anharmonic trapping potential on the scale of the dot ($8 \times 4$ \textmu{}m), the small extent of the electron ensemble ($0.5 \times 0.5$ \textmu{}m) allowed us to simulate the unloading with an axially symmetric harmonic well. The unloading voltage $V_\mathrm{unload}$ decreased the trap depth and resulted in unloading of the dot. We modeled this process as a linear decrease in barrier height: $V_b = V_\mathrm{bar} + \beta V_\mathrm{unload}$, where $V_\mathrm{bar} = 22$ meV was obtained from electrostatic simulations and $\beta$ was determined from the final jump $(A/A_0)^2$ in Fig.~\ref{fig:fig3}b. The energies of the Wigner molecules were calculated analytically \citemeth{MKongPRE2002}, which resulted in the unloading voltages $V_{\mathrm{unload}}^{(N)}$:
\begin{align}
    &V_{\mathrm{unload}}^{(1)} = -\frac{V_\mathrm{bar}}{\beta} = -0.305\, \mathrm{V} \notag \\ 
    &V_{\mathrm{unload}}^{(2)} = V_{\mathrm{unload}}^{(1)} + \frac{3}{4} \frac{E_0}{\beta e} \notag \\
    &V_{\mathrm{unload}}^{(3)} = V_{\mathrm{unload}}^{(1)} + 1.31037 \frac{E_0}{\beta e} \notag \\
    &V_{\mathrm{unload}}^{(4)} = V_{\mathrm{unload}}^{(1)} + 1.83545 \frac{E_0}{\beta e} \notag
\end{align}
where
\begin{equation}
    E_0 = \left( \frac{m_e \omega_e^2 e^4}{2\left( 4 \pi \right)^2 \varepsilon_0^2\varepsilon_\mathrm{He}^2}  \right)^{\frac{1}{3}}
\end{equation}
and depends only on the trap curvature at the unloading point ($\omega_e$), electron mass ($m_e$) and other physical constants. Best agreement between model and experiment was found with an effective trap curvature $\omega_e/2\pi = 26$ GHz, which produces the red arrows in Fig.~\ref{fig:fig3}b. 

If the dot had initially contained five electrons, our model would have predicted an additional plateau starting at  $V_\mathrm{unload}^{(5)}$ = -0.127 V. Since we did not observe this plateau we concluded the trap was initially loaded with $N=4$ electrons.
\\
\\
\textbf{Modeling of Wigner molecule spectra.}
To accurately model Wigner molecule spectra, we needed a more sophisticated model of the electrostatic potential than a axially symmetric harmonic well. Instead, the electrostatic potential was approximated by
\begin{equation}
    E/e = \alpha_0(V_\mathrm{trap}) x^2  + \alpha_1(V_\mathrm{trap}) y^2 + \alpha_2 (V_\mathrm{trap}) y ^4. \label{eq:methods_minimal_model}
\end{equation}
Without a quartic term, the method described below predicts crossings for all Wigner molecules at equal $V_\mathrm{trap}$, which is inconsistent with experiment. Eq. \eqref{eq:methods_minimal_model} represents a model that reproduces the observed spectroscopy traces. The coefficients $\alpha_i$ were obtained by first fitting Eq. \eqref{eq:methods_minimal_model} to the electrostatic potential obtained via finite element modeling, and were then slightly adjusted to reproduce the spectroscopy traces from experiment, using the following method. 

For a particular trap voltage the Wigner molecule configurations were found through numerical minimization of the total energy, which included a small screening correction to the interaction energy due to the metal electrodes under the electrons. In addition, we neglected the kinetic term in the total energy, since at $T=25$ mK the kinetic energy is approximately three orders of magnitude smaller than the interaction energy. Next, using the electron positions as input, the cavity frequency shift and orbital frequencies were determined by solving the linearized equations of motion of the coupled cavity-electron system. We then took the strongest-coupled orbital frequency $\omega_e$ and calculated its effect on the resonator via
\begin{equation}
    \frac{A}{A_0} = \left| \frac{\sqrt{\kappa_1 \kappa_2}}{i (\kappa_1 + \kappa_2 +\kappa_\mathrm{int})/2 - \chi(\omega_0)} \right|, 
    \label{eq:methods_input_output_model}
\end{equation}
where $\kappa_{1,2,\mathrm{int}}$ represents the coupling through port 1 and 2 of the resonator and the internal loss rate, respectively. In addition, the susceptibility is given by
\begin{equation}
    \chi(\omega_0) = \frac{g^2}{(\omega_0 - \omega_e) + i \gamma}.
\end{equation}
$g/2\pi$ was fixed at 5 MHz (estimated from the resonator geometry, see Supplementary Information) and $\gamma/2\pi$ was adjusted to get good agreement for $N=1$. $\gamma$ was not further adjusted for $N>1$ Wigner molecules, since for those molecules all orbital modes stayed far detuned and the modeled traces only weakly depended on $\gamma$. With this method we obtained the resonator responses shown as solid black traces in Fig.~\ref{fig:fig3}c. 

We obtained better agreement between the data and model for one and two electrons, compared with three and four electrons. This can be attributed to the larger size of the three and four-electron Wigner molecules, since the approximation of the electrostatic potential in Eq. \eqref{eq:methods_minimal_model} only holds for small $x, y$. In addition, each Wigner molecule spectrum was averaged about 2000 times which blurs sharp features, such as the one in the modeled three-electron trace.

The anharmonicity of a single electron was estimated by treating the $y^4$ term in Eq. \eqref{eq:methods_minimal_model} as a perturbation to the harmonic oscillator Hamiltonian. We define the anharmonicity $\alpha$ as $\hbar \alpha = (E_2 - E_1) - (E_1 - E_0)$, where $E_n$ are the perturbed eigenenergies. Near the crossing with the resonator we find $\alpha_2 \approx 0.014$ $\mu$m$^{-4}$, leading to
\begin{equation}
    \frac{\alpha}{2\pi} = \frac{1}{2\pi} \frac{3 e \alpha_2 \hbar}{m_e^2 \omega_e^2} \approx 85\, \mathrm{MHz}.
\end{equation}
\\
\\
\textbf{Extracting single electron properties.}
To extract $g$ and $\gamma$ from the data in Fig.~\ref{fig:fig4}c, we used the same model for the resonator transmission as in Eq. \eqref{eq:methods_input_output_model}, which was based on input-output theory and assumed that one orbital mode coupled to the resonator. To fit the frequency shift vs. trap voltage, we needed to know $\omega_e$ as function of $V_\mathrm{trap}$. We used quadratic fits to a finite element model of the electrostatic potential, which accurately predicted the single-electron crossing voltage, to find the dependence of $\omega_e$ on $V_\mathrm{trap}$. For the data in Fig.~\ref{fig:fig4}c, this method predicted a sensitivity near the crossing of $\partial f_e/\partial V_\mathrm{trap} = 95$ GHz/V and also gives the top horizontal axis in Fig.~\ref{fig:fig4}c.

Since the measured frequency shift remained less than a linewidth, the phase ($\Delta \varphi$) was a direct measure of the cavity frequency shift and the conversion was made via $\Delta \varphi = \arctan \left( \Delta f_0 / \kappa_\mathrm{tot} \right) \approx \Delta f_0 / \kappa_\mathrm{tot}$, where $\kappa_\mathrm{tot} = \kappa_1 + \kappa_2 + \kappa_\mathrm{int}$. Using the simulated $\omega_e$ vs. $V_\mathrm{trap}$, we fit the measured cavity frequency shift to $\Delta f_0 = \Delta \varphi \kappa_\mathrm{tot}$, which gave the values listed in the main text. Quoted uncertainties were fit uncertainties.


\onecolumngrid
\newpage
\appendix
\renewcommand{\thefigure}{S\arabic{figure}}
\setcounter{figure}{0}
\renewcommand{\figurename}{Supplementary Figure}
\renewcommand{\tablename}{Supplementary Table}
\renewcommand\theequation{S\arabic{equation}}
\setcounter{equation}{0}
\numberwithin{equation}{section}

\section*{Supplementary Material}

\section{Microwave resonator design and measurements}
\subsection{Design of the differential microwave mode}
\begin{figure}[hbtp]
\centering
  \centering
  \includegraphics[width=.35\textwidth]{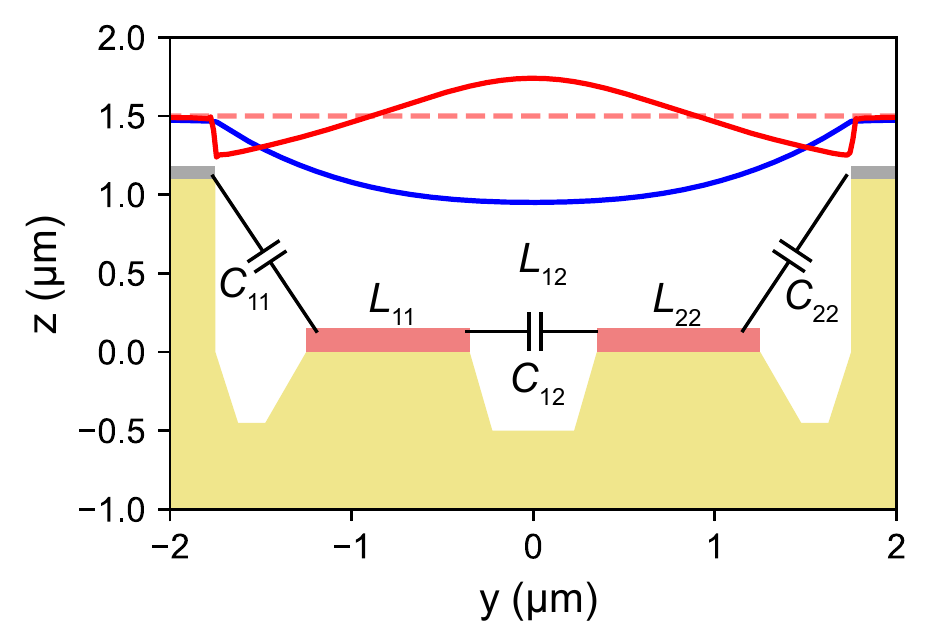}
  \caption{Schematic representation of the capacitances $C_{ij}$ and inductances $L_{ij}$ involved in a differential pair with two Nb center pins (red) and a ground plane (gray) on a Si substrate (yellow). The general shape of the DC potential (blue) and microwave electric field $E_y$ (red) are evaluated at the helium filling height $z$ = 1.2 $\mu$m. For the latter, a dashed line indicates $E_y = 0$.}
  \label{fig:differential_pair}
\end{figure}

To couple to the orbital electron state we use a superconducting microwave resonator consisting of two center pins surrounded by a ground plane. This geometry is schematically depicted in Fig. \ref{fig:differential_pair}. In general, this resonator geometry supports two types of modes. For the mode of interest, the pins carry an equal but opposite voltage at any point along the cross section. The microwave electric field is approximately constant between the two center pins, such that it can couple efficiently to the lateral motion of a single electron or a single row of electrons in the center of the micro-channel. 

The most essential microwave properties, e.g. the impedance $Z$ and resonance frequency $f_0$, can be extracted from the capacitances and inductances from Fig. \ref{fig:differential_pair}. The inductances and capacitances can be written in a matrix as follows: 
\begin{align}
	\mathcal{L} = 
\begin{pmatrix}
	L_{11}& L_{12}\\
	L_{21} & L_{22}
\end{pmatrix} \quad \text{and} \quad \mathcal{C} = 
\begin{pmatrix}
	C_{11}& C_{12}\\
	C_{21} & C_{22}
\end{pmatrix}.
\end{align}

Given our model geometry, each of the entries can be simulated using a finite element simulation package (e.g. Ansys Electronics Desktop). The impedance of the microwave differential mode is given by
\begin{align}
	Z_\text{diff} = 2\sqrt{\frac{L_{11} - L_{12}}{C_{11}+|C_{12}|}}.
    \label{eq:z_diff_resonator}
\end{align}
Without kinetic inductance (we estimate a kinetic inductance fraction of only 5\%) we estimate the characteristic impedance $Z_\mathrm{diff} \approx 90 \, \Omega$. Additionally, we find an expression for the expected resonance frequency for the quarter wavelength differential mode:
\begin{align}
	f_0 = \frac{1}{4 \ell} \frac{1}{\sqrt{(L_{11} - L_{12})(C_{11} + |C_{12}|)}},
\end{align}
where $\ell$ is the length of the resonator measured from the tip to the point where the two center pins meet.

\subsection{Electron-photon coupling} \label{app:single_electron_coupling_strength_estimate}
The coupling strength of a single electron to a single microwave photon can be estimated from the dipole energy
\begin{equation}
    \hbar g = \boldsymbol{d} \cdot \mathbf{E}. 
    \label{eq:coupling_strength_estimate}
\end{equation}
where $\boldsymbol{d} = e y_\mathrm{zpf} \mathbf{\hat{y}}$ is the dipole moment, $e$ is the electron charge and $y_\mathrm{zpf} = \sqrt{\hbar/2m_e \omega_e}$ is the zero point motion of the electron in the $\mathbf{\hat{y}}$-direction. The electric field $\mathbf{E} = E_y \mathbf{\hat{y}}$, where $E_\mathrm{y}$ is the electric field in the $y$-direction generated by the zero-point fluctuations of the microwave resonator $V_\mathrm{rms}$. The latter quantity can be estimated from the fact that on resonance half of the cavity's zero point energy is stored in the capacitor, such that
\begin{equation}
    \frac{1}{4} \hbar \omega_0 = \frac{1}{2}\frac{1}{T} \int_0^T \mathrm{d}t \cos^2(\omega_0 t) \int \mathrm{d}^3r  \varepsilon E^2(r) = \frac{1}{4} C V_\mathrm{zpf}^2 = \frac{1}{2} C V_\mathrm{rms}^2.
    \label{eq:voltage_fluctuations}
\end{equation}
This can be further simplified using the relations $\omega_0 = \sqrt{1/LC}$ and $Z = \sqrt{L/C}$, such that Eq. \eqref{eq:voltage_fluctuations} yields $V_\mathrm{rms} = \omega_0 \sqrt{\hbar Z/2}$. Plugging this in Eq. \eqref{eq:coupling_strength_estimate} gives 
\begin{equation}
    g/2\pi = \boldsymbol{d} \cdot \mathbf{E} = \frac{1}{2} e E_y f_0 \sqrt{\frac{Z}{m_e \omega_e}}.
\end{equation}
For realistic experimental values of $E_y \approx 2 \times 10^5$ V/m (see Fig. \ref{fig:dcbias_microwave_leakage}b), $Z = 90$ $\Omega$ and $f_0 = \omega_0/2\pi = 6.45$ GHz, we arrive at $g/2\pi \approx 5.0$ MHz.

\subsection{On-chip filter and microwave resonator characterization}

\begin{figure}[hbpt]
\centering
\includegraphics[width=0.5\columnwidth]{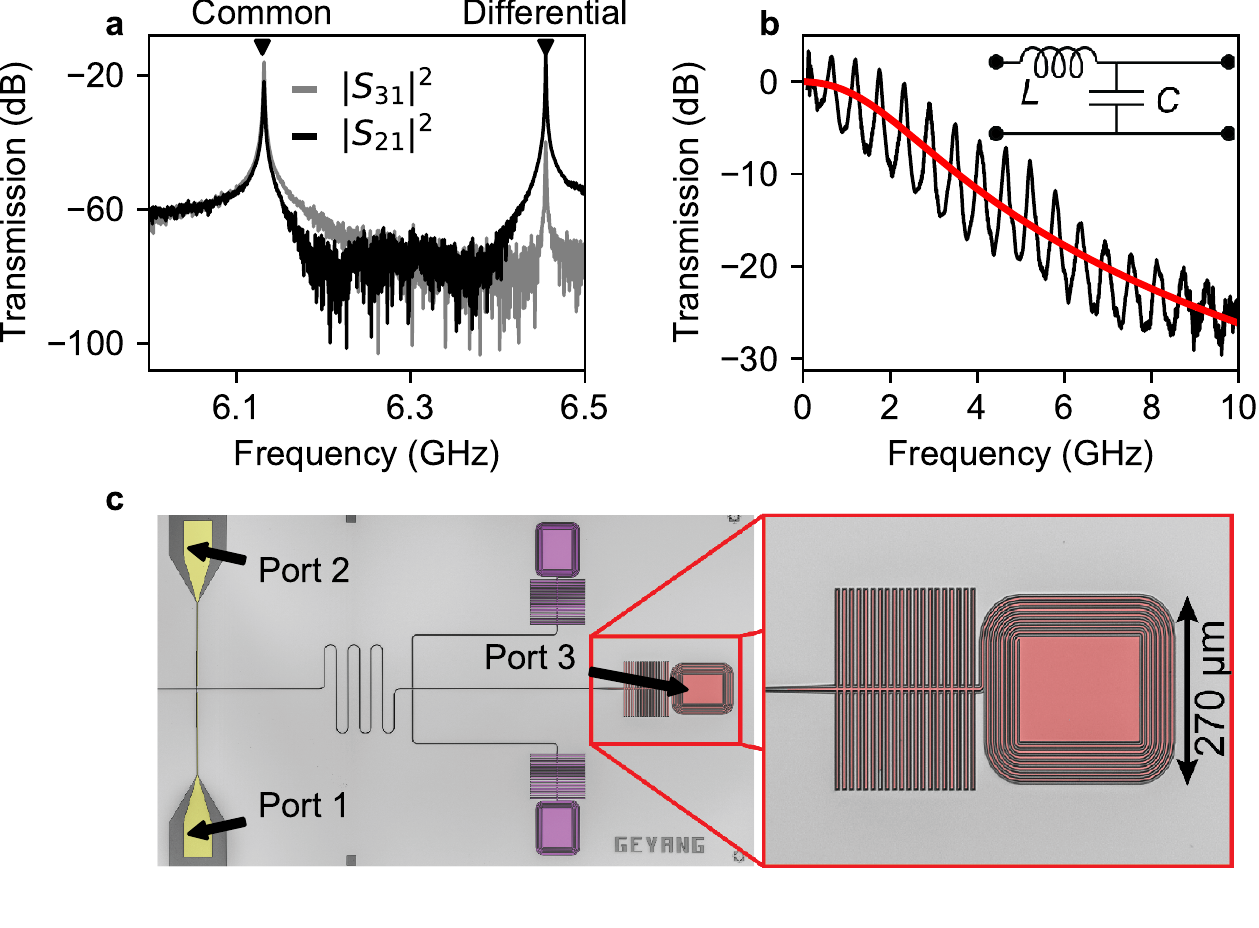}
\caption{(a) Microwave transmission measurements of the resonator, with ports labeled as shown in the inset. (b) Microwave transmission of the on-chip $LC$-filters can be modeled using a two port circuit model (red) as shown in the inset. (c) Optical micrograph of the spiral inductor that is part of the device's on-chip $LC$-filters. During the experiment these pads are flooded with helium and have to be positively biased, such that they accumulate unwanted electrons.}
\label{fig:supp_filter_char}
\end{figure}

To characterize the resonator we measure its transmission by driving and detecting through the yellow microwave feed lines in Fig. \ref{fig:supp_filter_char}c. The resulting transmission $|S_{21}|^2$ shows two peaks, separated by 300 MHz, which we identify as the common and differential mode of the microwave resonator. We identify the differential mode by also detecting the microwave transmission through the resonator DC bias line. Ideally, the transmission through the DC bias line is fully suppressed for the differential mode. However, due to asymmetry in the microwave field due to fabrication imperfections $|S_{31}|^2$ only shows a 27 dB reduced peak amplitude at 6.45 GHz. This is in contrast to the lower resonance at 6.15 GHz, for which the transmission amplitude increases. This indicates that the differential (common) mode has a resonance frequency of 6.45 (6.15) GHz. 

Without electrons or helium on top of the resonator, $S_{ij}$ is accurately described by
\begin{equation}
    S_{ij} = \frac{\sqrt{\kappa_i \kappa_j}}{(\omega-\omega_0) + i \kappa_\mathrm{tot}/2}, 
    \label{eq:supp_eq_sij}
\end{equation}
where $\kappa_\mathrm{tot} = \kappa_1 + \kappa_2 + \kappa_3 + \kappa_\mathrm{int}$ is the total line width and $\kappa_{1-3}$ are the coupling rates through ports 1-3. From a fit to this model, and defining the loaded quality factor as $Q_L = \omega_0 / \kappa_\mathrm{tot}$, we find $Q_L \approx 18 \times 10^3$ for the differential mode. Additionally, Eq. \eqref{eq:supp_eq_sij} allows us to estimate $\kappa_3$, since if ports 1 and 2 have equal coupling ($\kappa_1 = \kappa_2 = \kappa_c$), then
\begin{equation}
    \frac{\kappa_3}{\kappa_c} = \frac{|S_{31}(\omega = \omega_0)|^2}{|S_{21} (\omega = \omega_0)|^2}.
\end{equation}
Therefore, the 27 dB reduced transmission amplitude indicates that $\kappa_3/\kappa_c = 2 \times 10^{-3}$ and we conclude that $Q_L$ is not limited by leakage through the DC bias port.

We also measure the transmission of the on-chip microwave filters, using a separately fabricated chip containing just the filters and a through line for calibration of the amplitude. The result of this measurement is shown in Fig.  \ref{fig:supp_filter_char}b. The response shows a standing wave pattern, most likely due to an impedance mismatch between the chip and the printed circuit board. Apart from the oscillations, the overall response can be modeled well by a two-port $LC$-circuit as shown in the inset. We find that a capacitance of 4 pF and an inductance of 2.5 nH describe the response well (as shown by the red line). At the differential mode resonance frequency, the reflection is found to be 18 dB.

\subsection{Experimental setup}

\begin{figure}[hbtp]
\centering
  \centering
  \includegraphics[width=\textwidth]{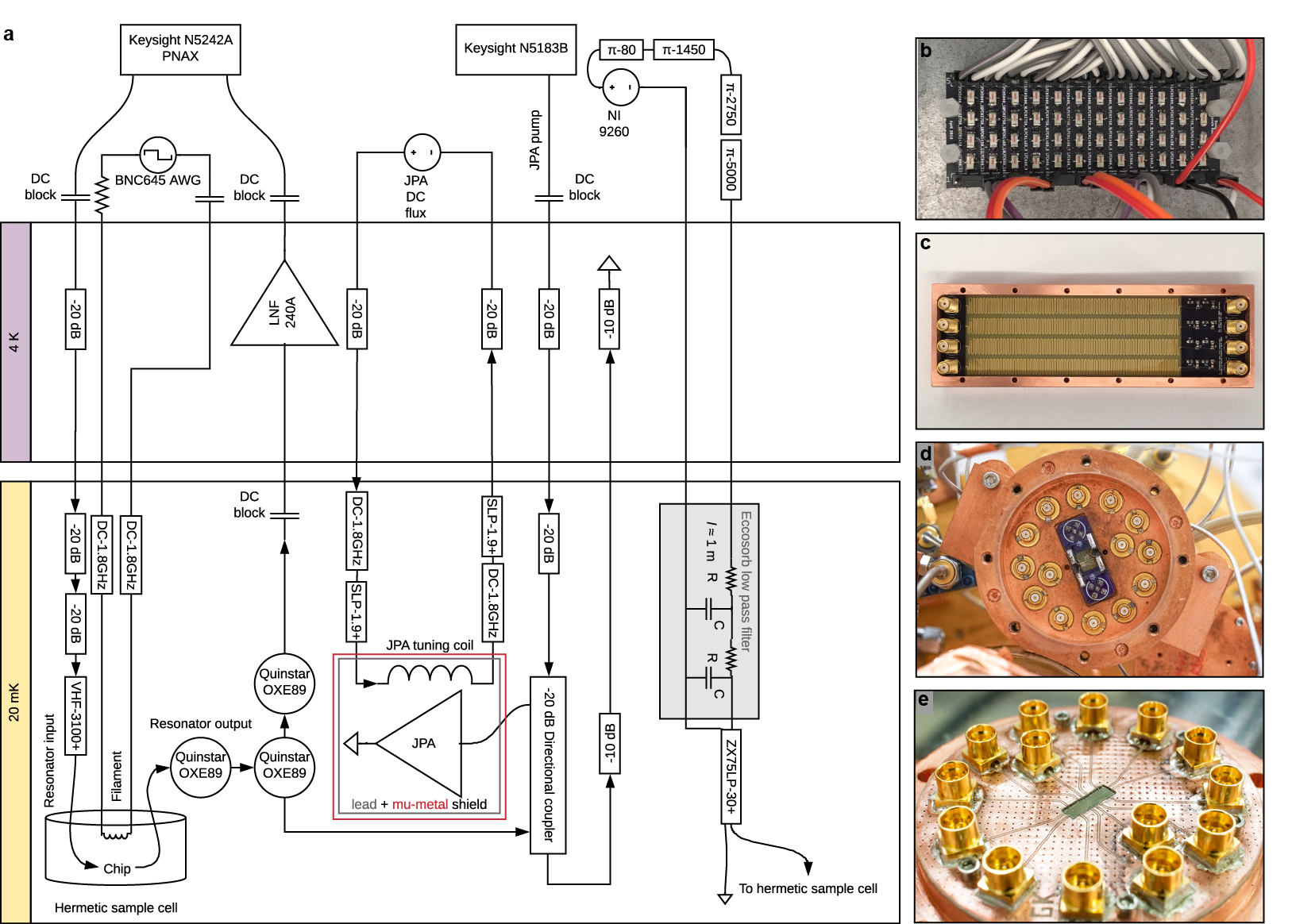}
  \caption{(a) Schematic representation of the experimental setup. Colored boxes represent different stages of the dilution refrigerator (50K, still and 100 mK plate not shown). Photographs of setup details are shown on the right. (b) Minicircuits $\pi$-filters (labeled as $\pi$-80, $\pi$-1450, $\pi$-2750 and $\pi$-5000 in (a)) mounted on a PCB at room temperature protect each pair of DC wires before entering the cryostat (c) copper enclosure containing RC filters and meandering wire (Eccosorb not shown) (d) Photograph of the inside of the sample box lid, showing 14 + 2 hermetic SMP connections which carry DC and microwave connections into the sample cell. The PCB in the center contains two tungsten filaments which are used as an electron source. (e) PCB with chip mounted to the bottom of sample cell. This part attaches to the lid shown in (d) and is sealed using a 0.020" OD indium seal applied around the circumference of the pedestal.}
  \label{fig:supp_exp_setup}
\end{figure}

Fig. \ref{fig:supp_exp_setup} shows a schematic diagram of the setup, not including the gas handling system that supplies helium to the sample box. These details can be found in the supplement of Ref. \citesm{SMGeYang2016}. 

\subsubsection{Microwave setup}
All microwave measurements were done with a Keysight PNA-X Network Analyzer. The transmitted signal from the microwave resonator is amplified by a Josephson parametric amplifier which provides a gain of approximately 20 dB at the cavity resonance frequency. The signal is subsequently amplified by a high electron mobility amplifier at 4 K (Low Noise Factory LNF-LNC48C, gain 38 dB) and a room temperature amplifier (Miteq AFS3-00101200, gain 28 dB). In addition, DC blocks (Inmet 8039) inserted in the in and output lines prevent ground loops.

\subsubsection{DC filtering}
Each DC electrode is low-pass filtered using a three stage filter attached to the mixing chamber plate of the refrigerator. The first two stages are combined on a custom designed printed circuit board, which is situated in a copper enclosure filled with Eccosorb CR117. The PCB contains pairs of long meandering traces \citesm{SMMuellerAPL2013} to increase the effective contact length with the lossy ferrite, and R-C filters with cut-off frequencies in the range 2-200 Hz. The third stage consists of a Minicircuits ZX75LP-30+ low pass filter that attenuates noise in the 30-3000 MHz range.

Additionally, an on-chip LC filter ($L \approx $ 2.5 nH, $C \approx$ 4 pF) for each electrode (attenuation of 18 dB at 6.5 GHz) further reduces the number of high frequency thermal photons that would otherwise degrade the cavity quality factor or adversely affect the electron motional state. \citesm{SMMi2016APL}

\subsection{Resonator response to superfluid helium}
As liquid helium fills the cylindrical reservoir below the chip, capillary action causes the channels to fill with liquid helium. As helium is added to the reservoir, the distance of the helium to the chip $h$ decreases and the micro-channel fills according to Jurin's law:
\begin{equation}
	\rho g h = \frac{\sigma}{R}, 
\end{equation}
where $\rho = 145$ kg/m$^3$ is the density of liquid helium, $\sigma=3.78\cdot10^{-4}$N/m$^2$ is the surface tension and $R$ is the radius of curvature of the helium-vacuum interface. To first order this equation states that the surface of the helium assumes the shape of a quadratic form $z(x, h)$:
\begin{equation}
	z(x, h) = d_0 + \frac{\rho g h}{2 \sigma} \left( x^2 - \frac{w^2}{4}  \right),
	\label{eq:quadratic_form}
\end{equation}
where $d_0=1.2\mu$m and $w=3.5\mu$m are the depth and width of the channel, respectively. The level of the liquid in the center of the channel is
\begin{equation}
	z(0, h) = t_\mathrm{He} = \max \left(0,  d_0 - \frac{\rho g h}{2 \sigma} \frac{w^2}{4} \right).
	\label{eq:supp_he_depth}
\end{equation}

\begin{figure}[hbtp]
\centering
  \centering
  \includegraphics[width=0.60\textwidth]{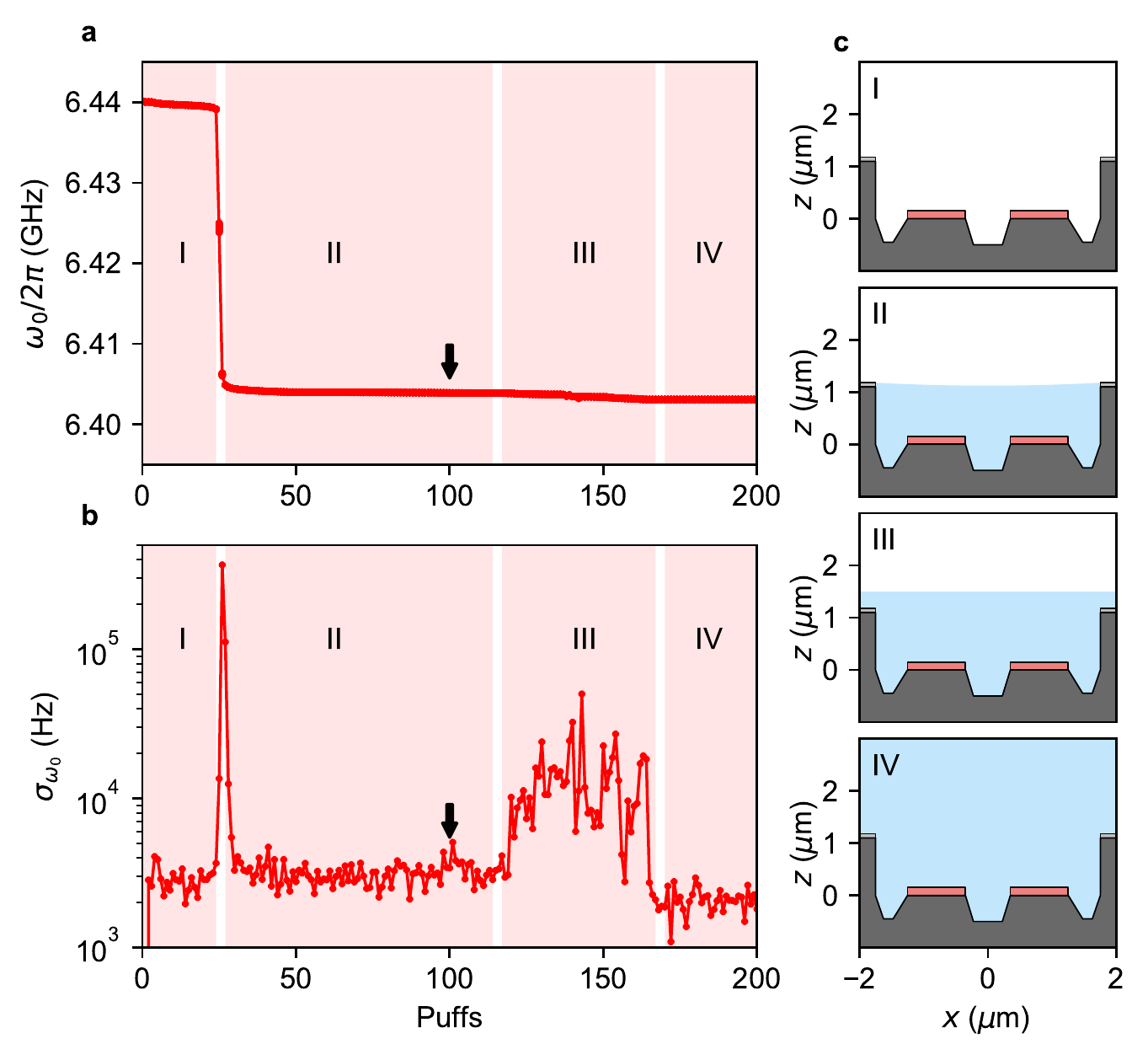}
  \caption{Resonator response to adding liquid helium to the sample cell. (a) Resonance frequency shift $\Delta \omega_0/2\pi$ and (b) Resonance frequency jitter $\sigma_{\omega_0}$ as function of the number of $^4$He gas puffs introduced to the sample cell. One puff corresponds to approximately 25 cc of $^4$He gas at STP. The experiment is performed with 100 puffs in the sample cell (black arrow). (c) Schematics of the channel showing the helium level (light blue) in each of the four regions denoted in (a) and (b).}
  \label{fig:supp_helium_level}
\end{figure}

The resonator frequency shift due to helium is depicted in Fig. \ref{fig:supp_helium_level}, where four different regions can be identified. In region I a $\sim$30 nm Vanderwaals film covers the entire sample, resulting in only a small frequency shift. Region II is characterized by a large jump in $\Delta \omega_0$ followed by a plateau. In this region helium fills the channel due to capillary action, until $h=0$. The plateau in $\omega_0/2\pi$ from 30-110 puffs can be explained by the channel geometry and the maximum value of $h$ set by helium reservoir depth. From Eq. \eqref{eq:supp_he_depth}, we estimate the helium depth in the plateau to vary from 1.0 to 1.2 $\mu$m. Introducing more helium results in filling the entire upper half plane (region III). The resonance frequency shift increases until the helium has filled the mode volume of the resonator. Beyond this point the electric field is negligible and, therefore, adding more helium does not result in an extra frequency shift (region IV).

At each point along the curve of Fig. \ref{fig:supp_helium_level}a, we repeatedly measure the resonance frequency $\omega_0$. The spread in $\omega_0$ at a particular helium filling is a result of superfluid helium vibrations that originate from continuous excitation from the pulse tube, and building vibrations that couple into the cryostat through its frame. In Fig. \ref{fig:supp_helium_level}b we plot the standard deviation $\sigma_{\omega_0}$ of 25 measurements of $\omega_0$. Note that each measurement of $\omega_0$ was acquired faster than the dominant frequency in the helium vibrational spectrum, such that the peak was not artificially broadened. Therefore, $\sigma_{\omega_0}$ gives a direct indication of the helium vibrations on the resonator.

Fig \ref{fig:supp_helium_level}b shows that helium vibrations are worst at the transition from region I to region II, i.e. just before the channel fills up with helium. In region II, the capillary action stabilizes the helium film and suppresses vibrations. An additional increase in helium vibrations is seen when the channel is completely full and capillary action no longer stabilizes the film.

\begin{figure}[hbtp]
\centering
  \centering
  \includegraphics[width=\textwidth]{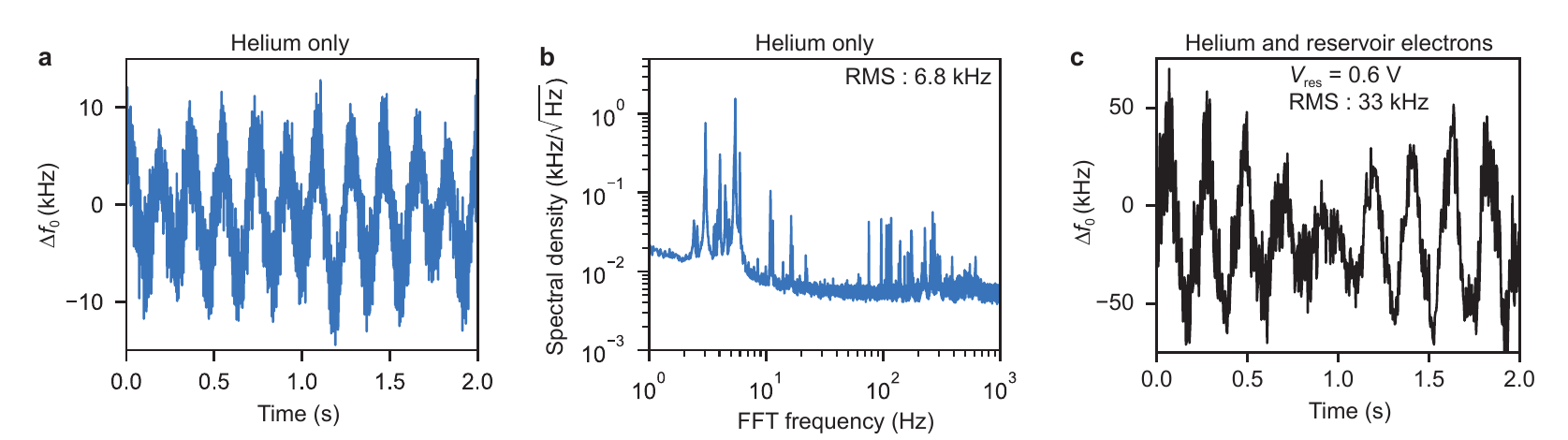}
  \caption{(a) Resonance frequency jitter due to helium vibrations measured at $T=25$ mK without reservoir electrons. (b) Most of the spectral density for the data in (a) lies below 10 Hz. Nearly all of the resonance frequencies in this region can be associated with a multiple of 1.4 Hz, the frequency of the pulse tube refrigerator. The quality factor of these modes are at least a few hundred. (c) After depositing reservoir electrons the frequency jitter increases. Note the difference in scale compared with (a). All time traces are taken with a microwave tone on resonance and converting the phase fluctuations to resonance frequency fluctuations using the resonator linewidth $\kappa_\mathrm{tot}$.}
  \label{fig:res_freq_jitter_w_electrons}
\end{figure}

Since helium vibrations are detrimental to the coherence of the electron orbital state, we decide to work at a point where $\sigma_{\omega_0}$ is at a minimum. The black arrow in Fig. \ref{fig:supp_helium_level}b shows this point. To further quantify the helium vibrations at this point, we monitor the resonance frequency as function of time and observe periodic oscillations with dominant frequencies less than 10 Hz (Fig. \ref{fig:res_freq_jitter_w_electrons}a,b). From the helium-resonator coupling (5 kHz/nm) we estimate the magnitude of classical helium fluctuations to be $\Delta t_\mathrm{He} = 1.4$ nm. The resonator frequency fluctuations due to these vibrations increases by a factor of five when reservoir electrons are present (Fig. \ref{fig:res_freq_jitter_w_electrons}c) because electrons couple more strongly to the resonator than helium.

Even though the magnitude of the jitter is less than a resonator linewidth $\kappa_\mathrm{tot}$, the variation of the resonance frequency shift over time obscures small frequency shifts due to electrons near the dot. We have tried various ways to minimize this noise, including turning off the pulse tube and working at elevated temperature to reduce the quality factor of the surface vibrations. Unfortunately, we see no improvement with the pulse tube turned off until $T > 0.3$ K and working at these temperatures introduces thermal noise which degrades the electron motional state noticably (see Appendix \ref{app:increased_temperature}). To circumvent the issue of the resonator jitter, we sweep the trap or guard voltages at a rate much faster than the dominant helium vibration frequency, such that frequency shifts from electrons in the dot become quickly apparent after averaging. A more quantitative description of the effect of helium vibrations on trapped electrons is given in Section \ref{sec:app_helium_vibrations_near_trap}.

\section{Comparison between experimental and modeled unloading voltages}

\begin{table}[hbtp]
\caption{Modeled unloading voltages compared with experimental jump locations.}
\begin{tabular}{cccl}
\hline
\hline
 Unloading voltage & Model (V) & Experiment (V) & Comment\\
 \hline
 \hline
 $V_{\mathrm{tg}}^{(1)}$& -0.305 & -0.305 & Taken from experiment and used as parameter in the model \\
 $V_{\mathrm{tg}}^{(2)}$& -0.248 & -0.246 & Obtained from fit with $\omega_e/2\pi = 26$ GHz\\
 $V_{\mathrm{tg}}^{(3)}$& -0.205 & -0.202 & Obtained from fit with $\omega_e/2\pi = 26$ GHz\\
 $V_{\mathrm{tg}}^{(4)}$& -0.165 & -0.168 & Obtained from fit with $\omega_e/2\pi = 26$ GHz\\
 $V_{\mathrm{tg}}^{(5)}$& -0.127 & & Not observed in experiment \\
 $V_{\mathrm{tg}}^{(6)}$& -0.092 & & Not observed in experiment \\
 \hline
\end{tabular}
\label{tab:unloading_theory_experiment}
\end{table}

\newpage
\section{Wigner molecule orbital frequencies} \label{sec:twotrappedelectrons}
\begin{figure}[hbtp]
\centering
  \centering
  \includegraphics[width=\textwidth]{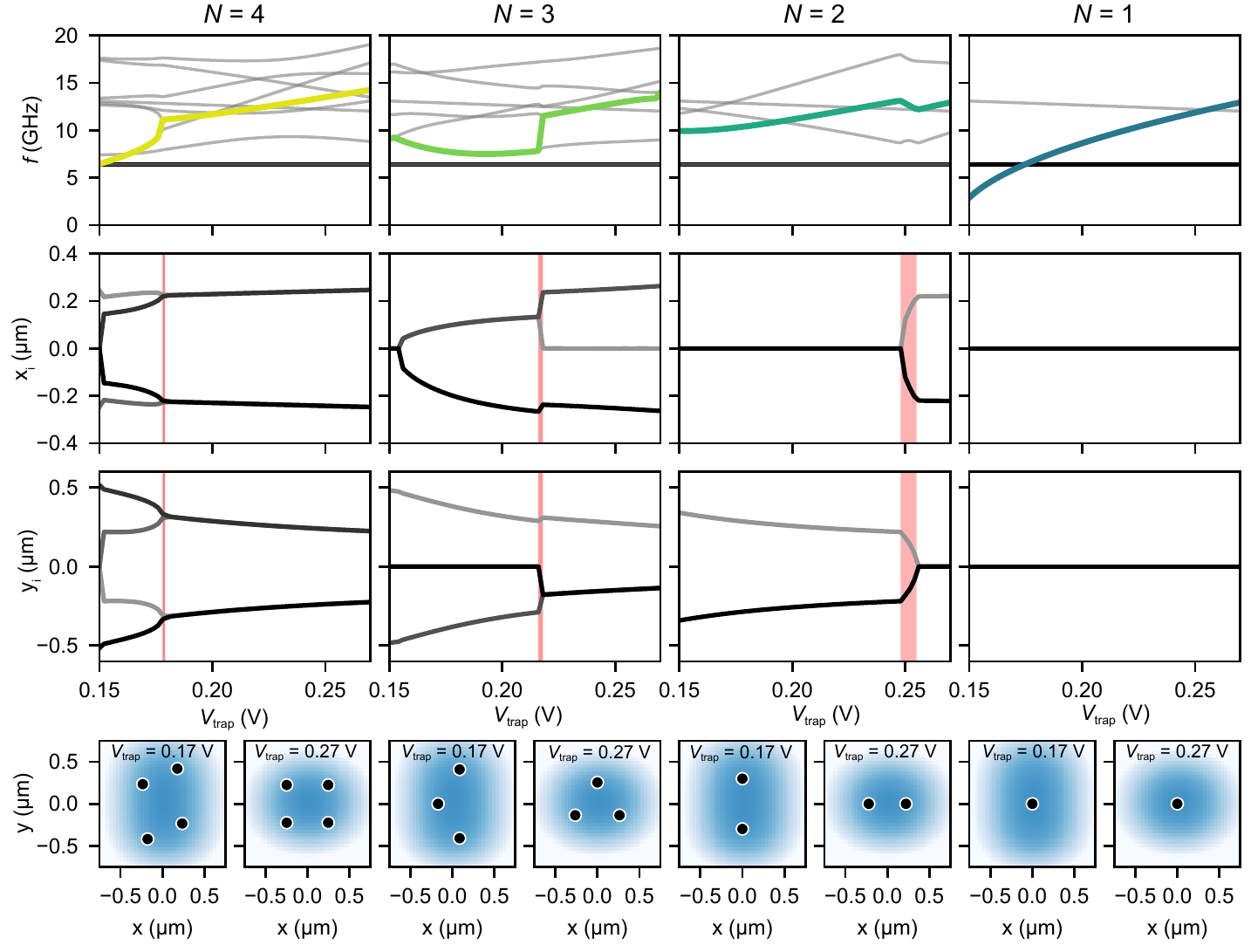}
  \caption{Mode frequencies and electron coordinates as function of $V_\mathrm{trap}$, associated with the solid black lines from Fig. 3c of the main text. In the top row, the strongest coupled electron mode is highlighted and the cavity mode is shown in black. Modes that couple weakly are shown in gray. The two center rows show electron rearrangements within small Wigner molecules at voltages indicated in red. For each $N$, snapshots of the electron configuration at $V_\mathrm{trap} = 0.17$ and 0.27 V further illustrate these rearrangements. The potential is shown in shades of blue, with the same colorbar as in Fig. 3d of the main text.}
  \label{fig:supp_mode_frequencies}
\end{figure}

For the coefficients $\alpha_i$ that reproduce the data of Fig. 3c, we plot the eigenfrequencies and electron positions ($x_i, y_i$) as function of $V_\mathrm{trap}$ in Fig. \ref{fig:supp_mode_frequencies}. Only for $N = 1$ the model predicts a crossing of an electron mode with the resonator. For higher $N$ none of the electron mode frequencies cross $f_0$ over the entire range of simulated $V_\mathrm{trap}$. Additionally, for $N=4, 3$ and 2, jumps in frequency and position indicate Wigner molecule rearrangements initiated by changes in the trap shape. For example, at low $V_\mathrm{trap}$ two electrons arrange in the across channel direction, whereas at higher $V_\mathrm{trap}$ it is energetically favorable to arrange in the along-channel direction. 

\newpage
\section{Simulation of anharmonicity of a single electron} \label{appendix:schrodinger_anharmonicity}
To use a single electron on helium as a qubit, its electrostatic potential needs to be anharmonic. To quantify the anharmonicity of the potential, we solve the Schr\"odinger equation for a single electron in a two-dimensional electrostatic potential where the spacing of the eigenstates reveals the anharmonicity. 

\begin{figure}[hbtp]
\centering
  \centering
  \includegraphics[width=.63\textwidth]{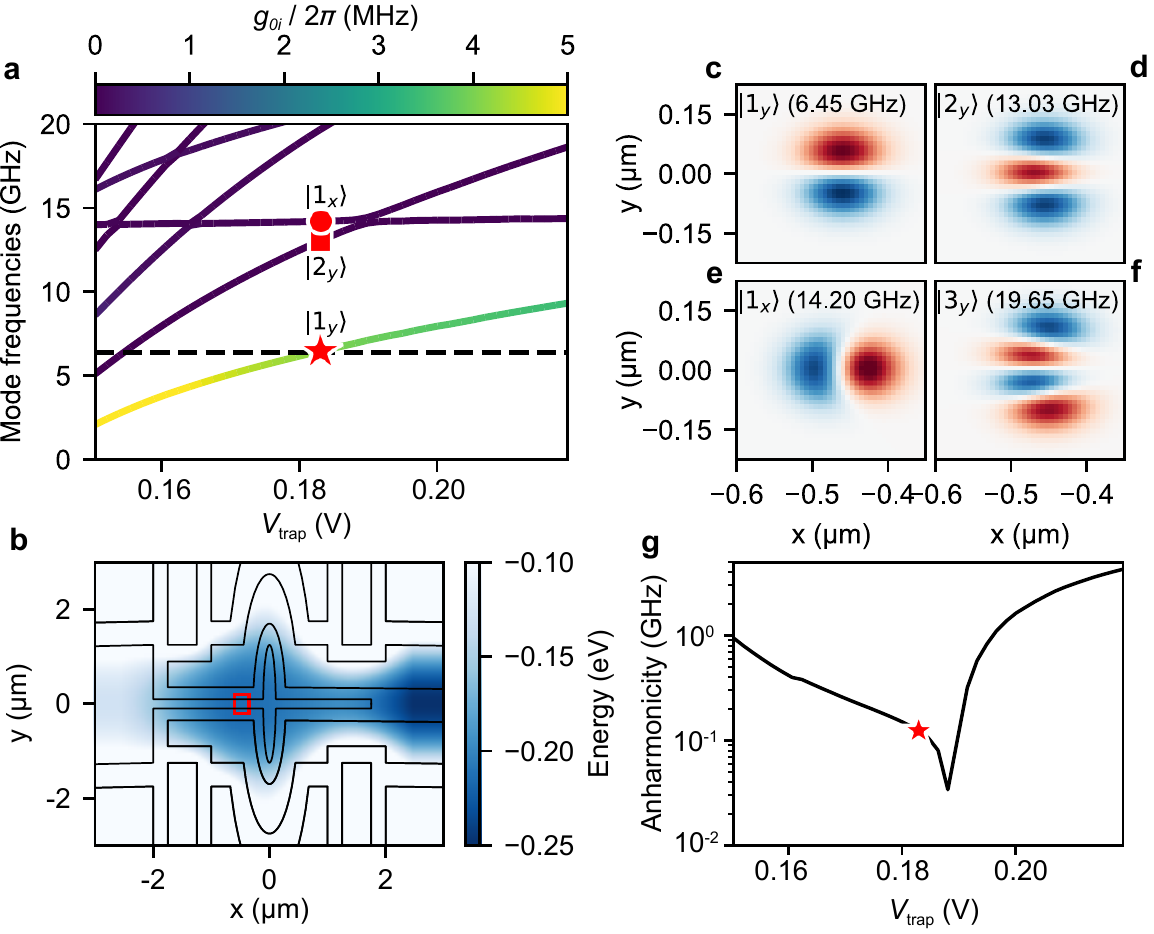}
  \caption{Quantum mechanical calculation of a single electron on helium (a) Transition frequencies for a single electron in the ground state, calculated by solving the Schr\"odinger equation in a two-dimensional electrostatic potential. The the coupling strength $g_{0i}$ for each state $|i\rangle$ is reflected in the color of each line. The resonator frequency is shown as a black dashed line. (b) Simulated electrostatic potential at $z = 1.15$ \textmu{}m and $V_\mathrm{trap}=0.184$ V. A red rectangle shows the extent of the single-electron wavefunctions shown in (c)-(f). (c)-(f) Wave functions of the excited states of a single electron near the crossing voltage $V_\mathrm{trap}$ = 0.184 V. (g) Inferred anharmonicity of a single electron as function of trap voltage. A red star marks the anharmonicity at the crossing voltage.}
  \label{fig:supp_wavefunctions}
\end{figure}

In Fig. \ref{fig:supp_wavefunctions}a we plot the transition frequencies from the ground state for the exact same electrode voltages as in Fig. 4 of the main text. The color of each line reflects the calculated coupling strength of each transition, which is calculated from the differential mode amplitude $V_\mathrm{RF}$ and the ground and excited state wavefunctions. Mathematically it takes on the form
\begin{equation}
	g_{0i}/2\pi = \frac{e}{2\pi} \iint \langle 0 | \left( x \frac{\partial V_\mathrm{RF}}{\partial x} + y \frac{\partial V_\mathrm{RF}}{\partial y} \right) |i \rangle dx \, dy,
\end{equation}
where $i = 0, 1_y, 1_x, 2_y, \ldots$ are the eigenmodes. It is clear that the ground state $|0\rangle$ is most strongly coupled to the first excited state in the $y$-direction (i.e. $|1_y \rangle$) and the coupling strength reaches several MHz, which is in agreement with the estimate from Appendix \ref{app:single_electron_coupling_strength_estimate}. Direct transitions from the ground state to other states are either forbidden by symmetry (e.g. $|0\rangle \Longleftrightarrow |2_y\rangle$) or extremely weakly coupled due to vanishing electric field (e.g. $|0\rangle \Longleftrightarrow|1_x\rangle$).

Fig. \ref{fig:supp_wavefunctions}a further correctly predicts a crossing of the $|0\rangle \Longleftrightarrow |1_y\rangle$ transition with the resonator around $V_\mathrm{trap} \approx 0.18$ V. At the crossing, which is indicated by a red star, the sensitivity is $\partial f_e / \partial V_\mathrm{trap} \approx 95$ GHz/V, and the wavefunction of $|1_y\rangle$ is shown in Fig. \ref{fig:supp_wavefunctions}b. The next two higher excited states at the crossing voltage are marked with a square and circle (Fig. \ref{fig:supp_wavefunctions}c and d, respectively), and we identify those as $|2_y\rangle$ and $|1_x\rangle$. Note the similarity between the wave functions from Fig. \ref{fig:supp_wavefunctions}b-e and those of a two-dimensional harmonic oscillator. However, unlike a harmonic oscillator, closer inspection of the transition frequencies reveals that the frequency spacing is non-uniform. In Fig. \ref{fig:supp_wavefunctions}f we plot the anharmonicity, defined as the difference between $f_{|1_y\rangle \rightarrow |2_y\rangle}$ and $f_{|0\rangle \rightarrow |1_y\rangle}$. At the crossing the anharmonicity exceeds 0.1 GHz, indicating that the electron can be approximated as a two-level system.

\newpage
\section{Single electron response to increased temperature} \label{app:increased_temperature}
In many circuit QED experiments temperature is an important parameter which, for example, controls excess photon noise and qubit dephasing. Experiments therefore operate at temperatures such that $k_b T \ll h f_0$ where $f_0$ is the transition frequency of the resonator or qubit. A natural question is how an electron on helium responds to increased temperature.

\begin{figure}[hbpt]
\centering
\includegraphics[width=\textwidth]{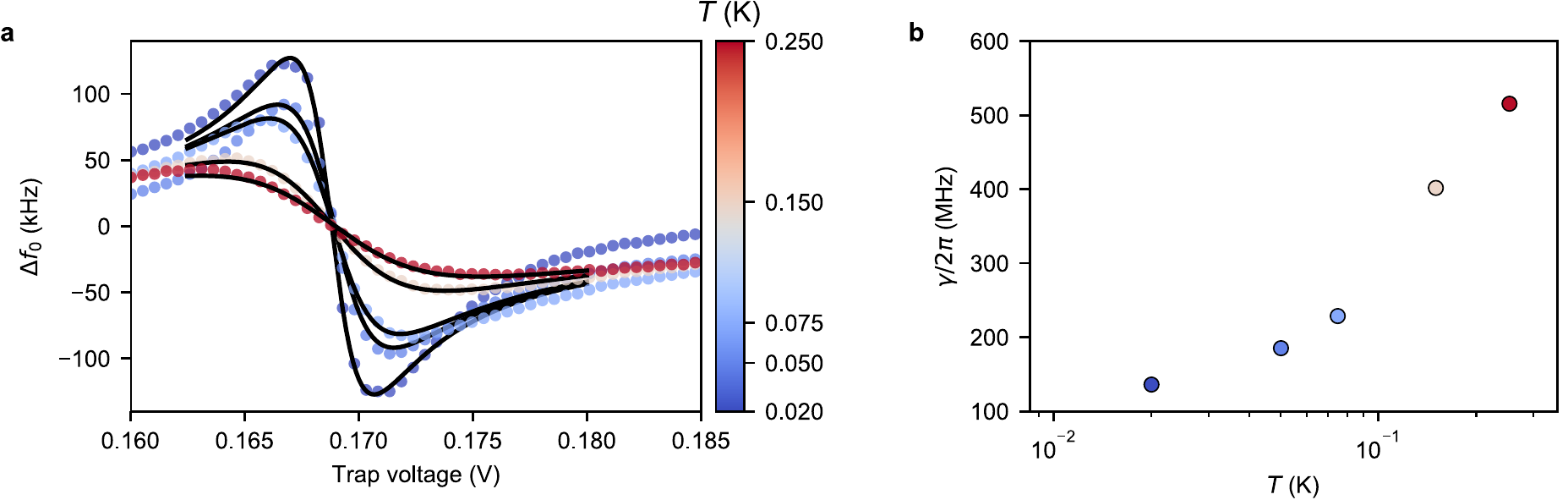}
\caption{(a) Single electron resonator spectroscopy traces as function of temperature, from cold (blue) to hot (red). (b) The extracted linewidths from fits to traces in (a) show an increase in linewidth of the orbital state.}
\label{fig:supp_temp_dependence}
\end{figure}

In Fig. \ref{fig:supp_temp_dependence} we plot the single electron resonator response as function of temperature. Since bias voltages are equal between traces, and the vibration amplitude of the superfluid helium surface is unaffected by temperature below $T \approx 0.25$ K, the observed broadening of the signal can only be attributed to heating of the electron. After fitting each trace (keeping $g$ constant between traces), we find that an increased temperature results in an increased linewidth, possibly due to thermal excitations of the orbital state. 

\newpage
\section{Contribution to single electron linewidth}
In this Appendix we discuss the possible noise sources that contribute to the measured single electron linewidth $\gamma$. In general, the linewidth can be written as the sum of the dephasing rate $\gamma_\varphi$, and transverse decay $\gamma_1$:
\begin{equation}
	\gamma = \frac{\gamma_1}{2} + \gamma_\varphi.
\end{equation}
An extensive list of decoherence mechanisms for the orbital state of an electron on helium is already available in the supplement of Ref. \citesm{SMSchuster2010}. Those calculations, which include the polarization of liquid helium, two ripplon decay processes, voltage noise through the electrodes and more, yield that $\gamma_1$ and $\gamma_\varphi$ should be sub-MHz. Since the observed linewidth is much larger, we consider additional sources of decoherence in the sections below. We list the magnitude of mechanisms and whether they contribute dephasing or decay in Supplementary Table \ref{tab:linewidth_contrib}. In the following section we briefly discuss each mechanism, starting with the dominant cause of dephasing: helium vibrations in the dot area.

\begin{table}[hbtp]
\caption{Summary of the contributions to the linewidth of a single electron from different types of noise or decay expected in our device.} 
\label{tab:linewidth_contrib}
\centering
\begin{tabular}{llc}
\hline
\hline
  Type & Mechanism & Magnitude \\
  \hline
  \hline
  Dephasing & Voltage noise from the gates &0.5 MHz\\
  Dephasing & Helium vibrations in the dot & 110 MHz\\
  Dephasing & Reservoir electrons on the resonator & 7 MHz \\
  Transverse & Microwave leakage through gates & $<$ 1 MHz \\
  \hline
\end{tabular}
\end{table}

\subsection{Helium vibrations in the dot}
\label{sec:app_helium_vibrations_near_trap}

\begin{figure}[hbtp]
\centering
  \centering
  \includegraphics{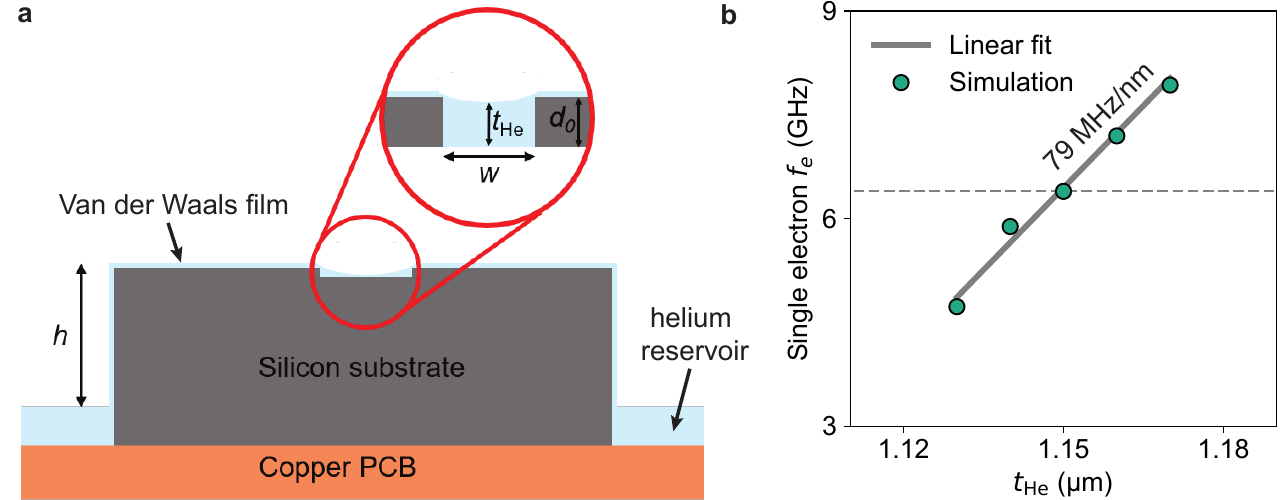}
  \caption{Helium vibrations and their effect on a single electron in the dot (a) Schematic of the helium in the channel and the helium in the off-chip reservoir, separated by a distance $h$. The zoom-in shows a close up of the helium inside the channel. (b) Frequency of a single electron trapped in the dot as function of the helium thickness measured in the center of the channel. For each $t_\mathrm{He}$, $f_e$ is determined at constant electrode voltages ($V_\mathrm{trap} \approx 0.184$ V). A linear fit gives the sensitivity of a single electron w.r.t. $t_\mathrm{He}$ near the resonator crossing ($f_e = f_0$), which contributes to the linewidth $\gamma$. The slope of this line is 79 MHz/nm.}
  \label{fig:supp_helium_reservoir}
\end{figure}
Since the electrostatic potential varies with the helium thickness $t_\mathrm{He}$, helium fluctuations in the dot are a source of dephasing. Helium thickness fluctuations in the micro-channel originate from vibrations in the reservoir, where the helium is not stabilized by surface tension. We can estimate how the magnitude of these vibrations scales with the channel geometry using Jurin's law. In the limit the channel is almost completely filled with helium (see Eq. \eqref{eq:quadratic_form}), the helium height in the center of the channel can be written as
\begin{equation}
	t_\mathrm{He} = d_0 - \frac{\rho g h w^2}{8 \sigma},
\end{equation}
where $h$ is the height from the chip to the reservoir level, $\sigma = 3.78 \times 10^{-4}$ N/m$^2$ is the surface tension of helium and $w$ is the channel width. Therefore, fluctuations in $t_\mathrm{He}$ due to level fluctuations inside the off-chip helium reservoir are given by 
\begin{equation}
	\Delta t_\mathrm{He} = \frac{\partial t_\mathrm{He}}{\partial h} \Delta h = \frac{\rho g w^2}{8 \sigma} \Delta h
    \label{eq:reservoir_flucs_sens}
\end{equation}
Eq. \eqref{eq:reservoir_flucs_sens} predicts that $\Delta t_\mathrm{He}$ scales as $w^2$, so helium fluctuations are expected to be worse near areas where the channel widens, such as the dot area and the spiral inductor.

From the measurements presented in Fig. \ref{fig:res_freq_jitter_w_electrons} and a simulated helium-resonator coupling of 5 kHz per nm of $^4$He, we estimate a magnitude of helium fluctuations of $\Delta t_\mathrm{He} \approx 1.4$ nm. With a single electron in the dot, the contribution from helium vibrations to the linewidth is then given by 
\begin{equation}
	\frac{\gamma_\varphi^\mathrm{He}}{2\pi} = \frac{\partial f_e}{\partial t_\mathrm{He}} \Delta t_\mathrm{He},
\end{equation}
where we estimate the electron sensitivity $\partial f_e/ \partial t_\mathrm{He} = 79$ MHz/nm (Fig. \ref{fig:supp_helium_reservoir}b). Finally, we arrive at the contribution due to helium fluctuations in the dot area: $\gamma_\varphi^\mathrm{He}/2\pi \approx 110$ MHz.

\subsection{Voltage noise from the gates}
Voltage noise on the electrodes in the dot area changes the electrostatic potential and thus leads to dephasing. It is either caused by electrical pickup or Johnson noise. We reduce it by using low-noise voltage sources and filtering the DC lines. Our $RC$-filters at the MC plate have a corner frequency of $f_\mathrm{3dB} \approx 300$ Hz for the resonator guard, trap guard and trap electrode and $f_\mathrm{3dB} \approx 4$ Hz for the resonator electrode. 

The linewidth due to voltage noise depends on the electron's sensitivity to each electrode, which we simulate by varying each voltage around the crossing voltage. The slope $\partial f_e/ \partial V_\mathrm{i}$ at the crossing voltage is a measure of the electron's sensitivity to electrode $i$. The noise on each electrode adds in quadrature, which leads to a total dephasing rate
\begin{equation}
	\frac{\gamma_\varphi^\mathrm{noise}}{2\pi} = \sqrt{\sum_i \left(\frac{\partial f_e}{\partial V_i} \right)^2 \Delta V_i^2}.
\end{equation}
Supplementary Table \ref{tab:linewidth_voltage_noise} lists sensitivities and the total dephasing rate, assuming each electrode has approximately 5 $\mu$V of voltage noise. The total estimated contribution due to voltage noise is 0.5 MHz.

\begin{table}[hbtp]
\caption{Simulated effect of each electrode on the single electron mode frequency $f_e$ and the resulting contribution to the single electron linewidth, assuming a voltage noise on each electrode of 5 $\mu$V.} \label{tab:linewidth_voltage_noise}
\centering
\begin{tabular}{lcc}
\hline
\hline
  Electrode $i$ & Simulated slope (GHz/V)& $\gamma_\varphi^i/2\pi$ (MHz) \\
  \hline
  \hline
  Resonator & -48 & 0.2\\
  Trap & 95 & 0.5\\
  Resonator guard & -7 & $<$0.1\\
  Trap guard & -11 & $<$0.1\\
  \hline
  Total&  & 0.5 \\
  \hline
\end{tabular}
\end{table}

\subsection{Helium vibrations on the resonator}
Reservoir electrons above the resonator form a capacitor with the image charges induced in the resonator electrode below. Fluctuations in the image charge in the resonator electrode affect a single electron in the dot since the resonator electrode also extends into the dot area and its lever arm is nonzero. The capacitance of the sheet of electrons on the resonator can be approximated by a parallel plate capacitance: $C = \varepsilon_0 \varepsilon_\mathrm{He} A / t_\mathrm{He}$, where $t_\mathrm{He}$ is the height of the electrons above the electrode, and $A$ the effective area of the sheet. The voltage drop across the two charge sheets is then given by 
\begin{equation}
	V = \frac{Q}{C} =  \frac{N}{A} \frac{e t_\mathrm{He}}{\varepsilon_0 \varepsilon_\mathrm{He}} = \frac{n e t_\mathrm{He}}{\varepsilon_0 \varepsilon_\mathrm{He}},
\end{equation}
where $n$ is the electron density. The voltage noise due to a fluctuating helium level is given by 
\begin{equation}
	\Delta V_\mathrm{res} = \frac{\partial V}{\partial t_\mathrm{He}} \Delta t_\mathrm{He} = \frac{n e}{\varepsilon_0 \varepsilon_\mathrm{He}} \Delta t_\mathrm{He}
    \label{eq:voltage_noise_reservoir_electrons}
\end{equation}
We estimate the dephasing from the electron sensitivity to the resonator electrode (see Supplementary Table \ref{tab:linewidth_voltage_noise}):
\begin{equation}
	\frac{\gamma_\varphi^\mathrm{res}}{2\pi} = \left| \frac{\partial f_e}{\partial V_\mathrm{res}} \right| \frac{n e}{\varepsilon_0 \varepsilon_\mathrm{He}} \Delta t_\mathrm{He}
\end{equation}
Again, assuming $\Delta t_\mathrm{He} \approx 2$ nm and a typical reservoir electron density of $4 \times 10^{12}$ m$^{-2}$, we estimate a voltage noise of 0.1 mV and $\gamma_\varphi^\mathrm{res}/2\pi \approx $ 7 MHz. We assume there are no electrons on the trap, resonator guards or trap guards, such that a similar calculation for these electrodes does not result in additional dephasing. In future devices, this source of dephasing can be eliminated completely by removing the reservoir electrons or using an additional reservoir that does not couple to the resonator. 

\subsection{Microwave leakage through DC bias electrodes}

\begin{figure}[hbtp]
\centering
\includegraphics{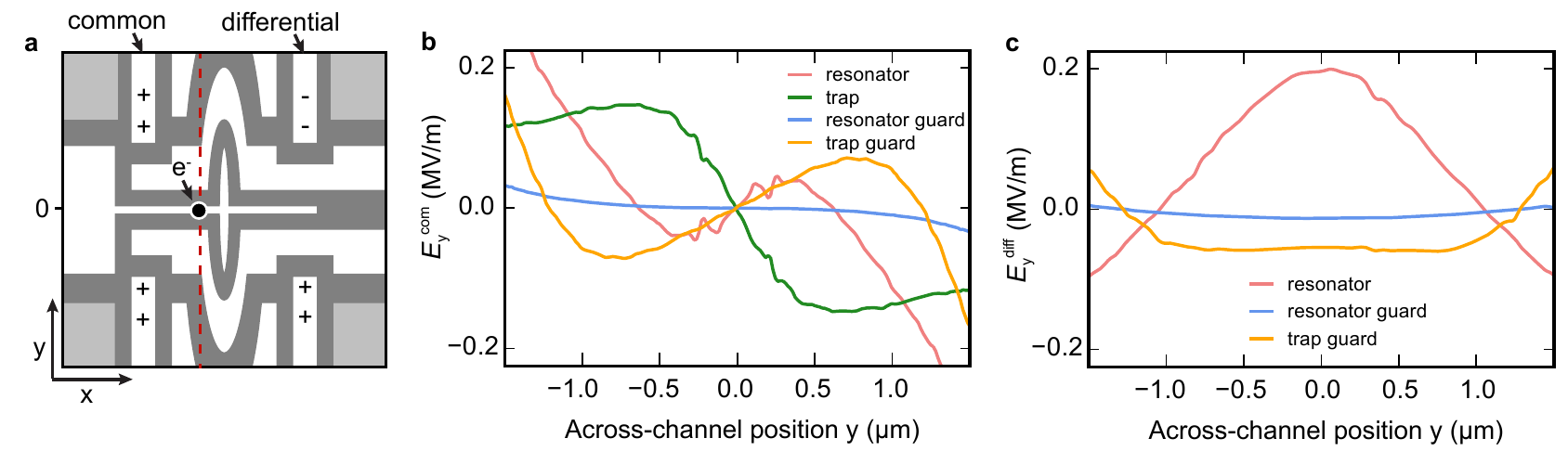}
\caption{(a) Schematic of the electrode geometry in the dot area, showing the expected electron position and an example of charge distributions on the trap guard and resonator guard which correspond to common and differential microwave leakage, respectively. (b) Common mode $E_y$ simulated at the red dashed line in (a). The values of $E_y$ at the electron position provide an estimate for microwave leakage through each respective electrode. (c) Same as in (b), but for the differential mode. Note that the trap electrode does not support a differential mode.}
\label{fig:dcbias_microwave_leakage}
\end{figure}

Ideally, the resonator is the only electrode that couples to the electron's motion and DC bias electrodes form perfect microwave reflectors from the electron's perspective. In practice there is always some leakage, even though we have taken the following measures to reduce this unwanted effect:
\begin{itemize}
    \item Adding a low-pass $LC$-filter on each DC bias electrode, and
    \item Shorting the left and right electrodes of each pair of guard electrodes. 
\end{itemize}
To quantify microwave leakage, we note that the coupling strength is set by $g = \mathbf{d}\cdot \mathbf{E}$. The electric field of each electrode in the across-channel direction therefore determines the leakage (i.e. coupling). Except for the trap electrode, we model microwave emission into each electrode by considering a common and differential mode, which are shown in  Fig. \ref{fig:dcbias_microwave_leakage}a, b. Since the electron couples to the differential mode of the resonator with $g/2\pi \approx 5$ MHz, and all other electric fields at the electron position are much smaller than $E_y^\mathrm{diff}$ of the resonator, we estimate the total decay from leakage through the bias electrodes to be $\gamma/2\pi < 1$ MHz.
\newpage
\section{Signs of helium vibrations in the crossing spectrum}
\begin{figure}[hbtp]
\centering
  \centering
  \includegraphics[width=.7\textwidth]{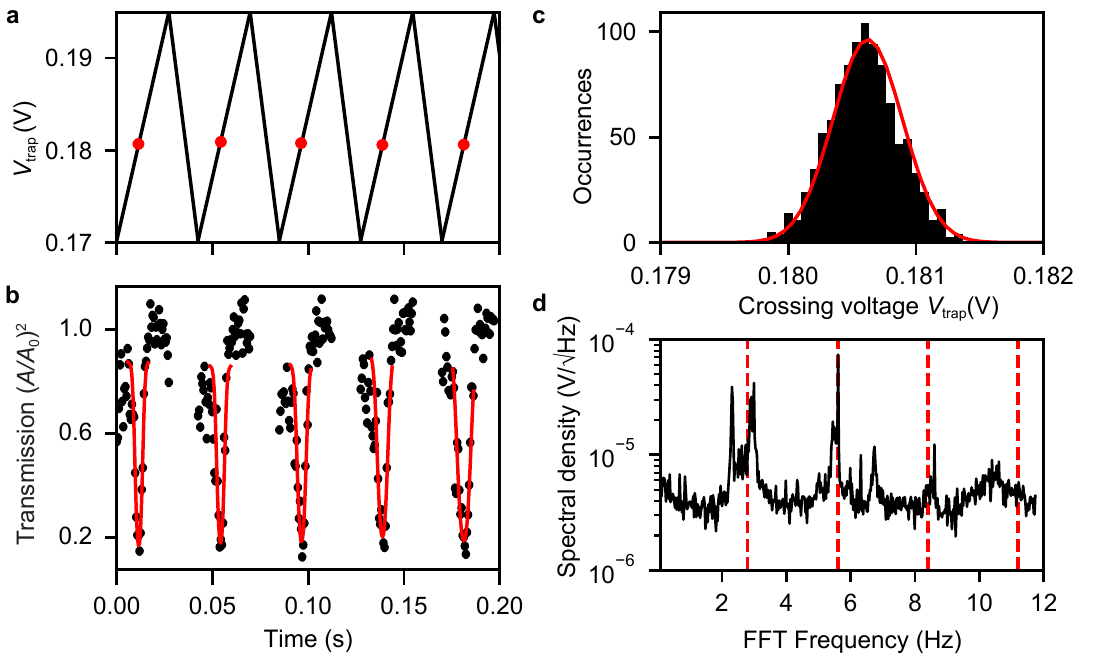}
  \caption{Statistics on fast single electron sweeps. (a) After loading a single electron we sweep the trap over a range of 25 mV while measuring the resonator transmission. A red dot marks the determined crossing voltage found by fitting the signal displayed in (b). (b) The resonator transmission shows unaveraged crossings of a single electron with the resonator (black dots). The solid red lines are fits which give the time and voltage of the crossing. (c) Statistics of $10^4$ crossings with the resonator. The solid red line is a Gaussian fit with standard deviation $\Delta V_\mathrm{trap} = 0.28$ mV. (d) Fourier transform of the crossing voltage time series. The dashed lines at 2.8, 5.6, 8.4 and 11.2 Hz indicate multiples of the pulse tube refrigerator frequency, and align with peaks in the crossing spectrum.}
  \label{fig:pulse_tube_in_electron}
\end{figure}

Here we present extra evidence to support our claim that the single-electron linewidth is significantly affected by helium fluctuations. Since helium fluctuations change the single-electron orbital frequency, its crossing voltage with the resonator is expected to vary with time. We attempted to measure this effect by repeatedly bringing the orbital frequency into resonance using the trap electrode. The JPA ensures a high signal-to-noise ratio, such that we can accurately estimate the crossing voltage by fitting the normalized transmission during a single voltage ramp (see Fig. \ref{fig:pulse_tube_in_electron}b). After repeating the experiment $10^4$ times, we obtain an average crossing voltage of $V_\mathrm{trap} \approx 0.1805$ V with a standard deviation of $\Delta V_\mathrm{trap} = 0.3$ mV. Using the simulated sensitivity of the trap electrode of $\partial f_e / \partial V_\mathrm{trap} = 95$ GHz/V, this spread in the crossing voltage corresponds to a single electron linewidth
\begin{equation}
    \gamma/2\pi = 2 \sqrt{2\ln 2} \frac{\partial f_e}{\partial V_\mathrm{trap}} \Delta V_\mathrm{trap} = 67 \, \mathrm{MHz},
\end{equation}
which agrees with the value from the main text: $\gamma_\varphi /2\pi = (77 \pm 19)$ MHz.

Since the time of each crossing is known from the ramp, we can Fourier transform the crossing time series (red dots in Fig. \ref{fig:pulse_tube_in_electron}a) to learn about the spectral content. The spectrum of the crossing voltage shows distinct peaks at even multiples of the pulse tube refrigerator (1.4 Hz) and looks very similar to the bare helium fluctuation spectrum measured in Fig. \ref{fig:res_freq_jitter_w_electrons}b. Therefore, these data directly show the effect of helium vibrations on a single electron. 

We have attempted to refocus individual crossings from Fig. \ref{fig:res_freq_jitter_w_electrons}b in post-processing but did not observe an increase in linewidth after fitting the averaged refocused data. It is possible that helium vibrations with frequencies larger than 12 Hz still contribute significantly to the spectrum. The maximum frequency we can detect in the crossing spectrum is limited by the repetition rate of the experiment ($f_\mathrm{rep} \approx 1/45$ms). For this experiment the corner frequency of the $RC$-filters prevented measurement of higher frequency components in the spectrum. However, by removing these filters this technique could be used to characterize the spectral density of a single electron even at higher frequencies.

\end{document}